# Parametric Nonlinear Optics with Layered Materials and Related Heterostructures

*Oleg Dogadov, Chiara Trovatello, Baicheng Yao, Giancarlo Soavi,* and Giulio Cerullo**

**Nonlinear optics is of crucial importance in several fields of science and technology with applications in frequency conversion, entangled-photon generation, self-referencing of frequency combs, crystal characterization, sensing, and ultra-short light pulse generation and characterization. In recent years, layered materials and related heterostructures have attracted huge attention in this field, due to their huge nonlinear optical susceptibilities, their ease of integration on photonic platforms, and their 2D nature which relaxes the phase-matching constraints and thus offers a practically unlimited bandwidth for parametric nonlinear processes. In this review the most recent advances in this field, highlighting their importance and impact both for fundamental and technological aspects, are reported and explained, and an outlook on future research directions for nonlinear optics with atomically thin materials is provided.**

## 1. Introduction

Nonlinear optical (NLO) effects manifest themselves as the response of matter to the interaction with an intense electric field $E$, typically laser light. In this case, the induced polarization inside the material becomes a nonlinear function of the electric field of light, which in a simplified scalar form reads $P = \varepsilon_0[\chi^{(1)}E + \chi^{(2)}E^2 + \chi^{(3)}E^3 + \cdots]$, where $\varepsilon_0$ is the permittivity of free space (SI system). The $n$th order nonlinear susceptibilities $\chi^{(n)}$ mix the electric fields incident on the material, allowing a series of algebraic (sum and difference) operations with photon energies. Efficient nonlinear frequency conversion requires both energy and momentum conservation. For bulk samples, the latter condition, also known as phase matching, is typically obtained exploiting birefringence in anisotropic nonlinear crystals (e.g., beta-barium borate with $\chi^{(2)}$ values of a few pm $V^{-1}$), or with periodically-poled nonlinear crystals (e.g., lithium niobate with $\chi^{(2)} = 10-20$ pm $V^{-1}$).[1] In such nonlinear crystals there is a stringent compromise between phase-matching bandwidth and total conversion efficiency, which are respectively inversely and directly proportional to the material's thickness. In addition, their integration on photonic platforms requires complex and expensive fabrication steps. For these reasons, 2D materials have emerged in the last decade as ideal candidates for next-generation ultrafast and ultrathin integrated nonlinear optical devices. 2D materials have exceptionally large optical nonlinearity (10–100× higher than standard bulk crystals),[2] are easy to integrate on photonic platforms by different methods[3–6] and, due to their atomically thin nature, offer the ultimate platform for frequency mixing beyond phase-matching constraints.[7]

In recent years, graphene, transition metal dichalcogenides (TMDs), and their related heterostructures have been widely exploited as nonlinear materials for coherent frequency conversion at the nanoscale. The field of nonlinear optics with 2D materials has been rapidly evolving and still continues to show new exciting physics along with promising technologically relevant results. For a general overview of the topic, we refer the reader to some classical introductory books on nonlinear optics[8,9] and to previously published reviews on the basic nonlinear optical properties of 2D materials.[2,10–19] Instead, in this review we will focus on the most recent results, achieved in the last few years, related to parametric coherent nonlinear optical processes in 2D materials and related heterostructures, highlighting both their fundamental aspects and the possible cutting-edge technological applications, as schematically depicted in **Figure 1**. On the one hand, NLO studies shed light on the fundamental electronic, magnetic, and optical properties of 2D materials; on the other hand, the studies of NLO properties of 2D materials pave the way for advanced technological applications.

The paper is structured in eight main sections.

O. Dogadov, C. Trovatello[+], G. Cerullo
Dipartimento di Fisica
Politecnico di Milano
Piazza Leonardo da Vinci, 32, 20133 Milano, Italy
E-mail: giulio.cerullo@polimi.it

B. Yao
Key Laboratory of Optical Fiber Sensing and Communications (Education Ministry of China)
University of Electronic Science and Technology of China
611731 Chengdu, China

G. Soavi
Institute of Solid State Physics
Friedrich Schiller University Jena
Max-Wien Platz 1, 07743 Jena, Germany
E-mail: giancarlo.soavi@uni-jena.de

The ORCID identification number(s) for the author(s) of this article can be found under https://doi.org/10.1002/lpor.202100726

[+]Present address: Department of Mechanical Engineering, Columbia University, New York, NY 10027, USA

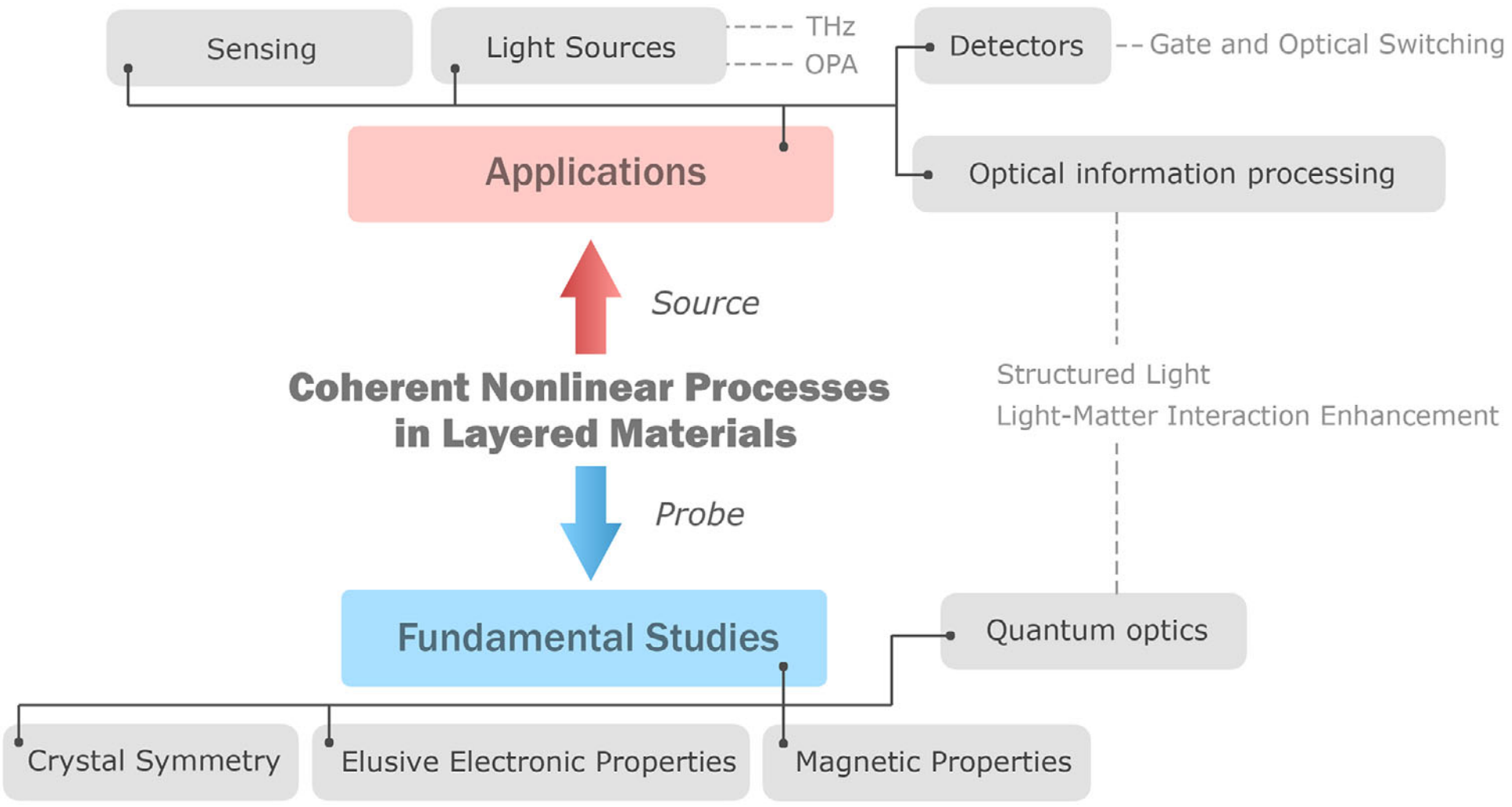

**Figure 1.** Flowchart of the topics covered in the review.

In Section 2 we discuss the fundamental aspects of the most common parametric NLO effects, that is, second order nonlinear processes (second harmonic generation, SHG, sum/difference frequency generation, SFG/DFG, and optical parametric amplification, OPA), third order nonlinear processes (third harmonic generation, THG, and four-wave mixing, FWM), high harmonic and THz harmonic generation (HHG). In Section 3 we discuss how NLO processes can be exploited to efficiently probe the crystal structure, the electronic properties (including the valley degree of freedom), interlayer charge transfer, quantum phases and phase transitions in 2D materials and related heterostructures. Finally, we also discuss the possibility to use NLO to explore exotic quantum interference pathways, which translate into light-induced transparency regimes.

Section 4 presents the newest results on electro-optical and all-optical modulation of NLO processes in 2D materials, and their promising future for next-generation ultrafast nonlinear optical devices.

Section 5 is dedicated to NLO-based sensing applications, that is, gas and molecule sensing.

In Section 6 we present the newest strategies for enhancing the efficiency of NLO processes in 2D materials, including metasurfaces, gratings, plasmonic enhancement, quantum dots (QD), nanowires (NW), fibers, waveguides, alternative non-centrosymmetric crystal structures, and hybridization with epsilon-near-zero (ENZ) materials.

Section 7 is dedicated to the generation of structured light using NLO, which offers the unique possibility to control the degree of freedom of the orbital angular momentum (OAM). Different from the spin, which is limited only to two possible states, OAM can assume an infinite number of possible values, thus offering new opportunities for quantum optics, imaging, and communications.

Finally, in the concluding Section 8, we give an outlook on the open challenges and future perspectives of NLO with 2D materials. We envision exciting opportunities for TMD-based on-chip parametric amplification, twistronics, that is, the possibility to exploit the interlayer twist angle in vertical 2D heterostructures to tune their NLO properties, and quantum optics, specifically for ultra-broadband entangled photon generation via spontaneous parametric down conversion (SPDC).

## 2. Frequency Conversion

In nonlinear optics, a process is defined as parametric if the initial and final quantum states are the same, so that population is only excited to virtual levels and energy is always conserved.[9] In this section we give an overview of the main parametric NLO processes (i.e., HHG, SFG and DFG, OPA and FWM) occurring in 2D materials. We also briefly discuss the perturbative regime of high harmonic generation in graphene and TMDs and the terahertz harmonic generation in graphene within a thermodynamic picture.

### 2.1. Second Order Nonlinear Optical Processes

In the dipole approximation, second order nonlinear optical processes such as SHG, SFG and DFG can only occur in materials with broken inversion symmetry.[9]

#### *2.1.1. Second Harmonic Generation*

In SHG two photons with energy $\hbar\omega$ are annihilated to create one photon with energy $2\hbar\omega$ (see the inset in **Figure 2**a). SHG was observed in a large number of layered materials, for example, $MoS_2$, $MoSe_2$, $WS_2$, $WSe_2$, $MoTe_2$, and hexagonal boron nitride (hBN), that belong to the $D_{3h}$ symmetry group for odd number of layers,[20–24] while it is symmetry forbidden in single-layer (1L) graphene and in TMDs with an even number of layers. Similar to semiconducting crystals, like gallium arsenide,[8] TMDs have huge $\chi^{(2)} = 100 - 1000$ pm V$^{-1}$, orders of magnitude higher than standard anisotropic nonlinear crystals, as it was mentioned in Section 1. Differently from bulk nonlinear crystals which have a large efficiency but macroscopic thickness, the unique advantage of TMDs is the possibility to be thinned down to nanometers and below, naturally lending to on-chip integration.

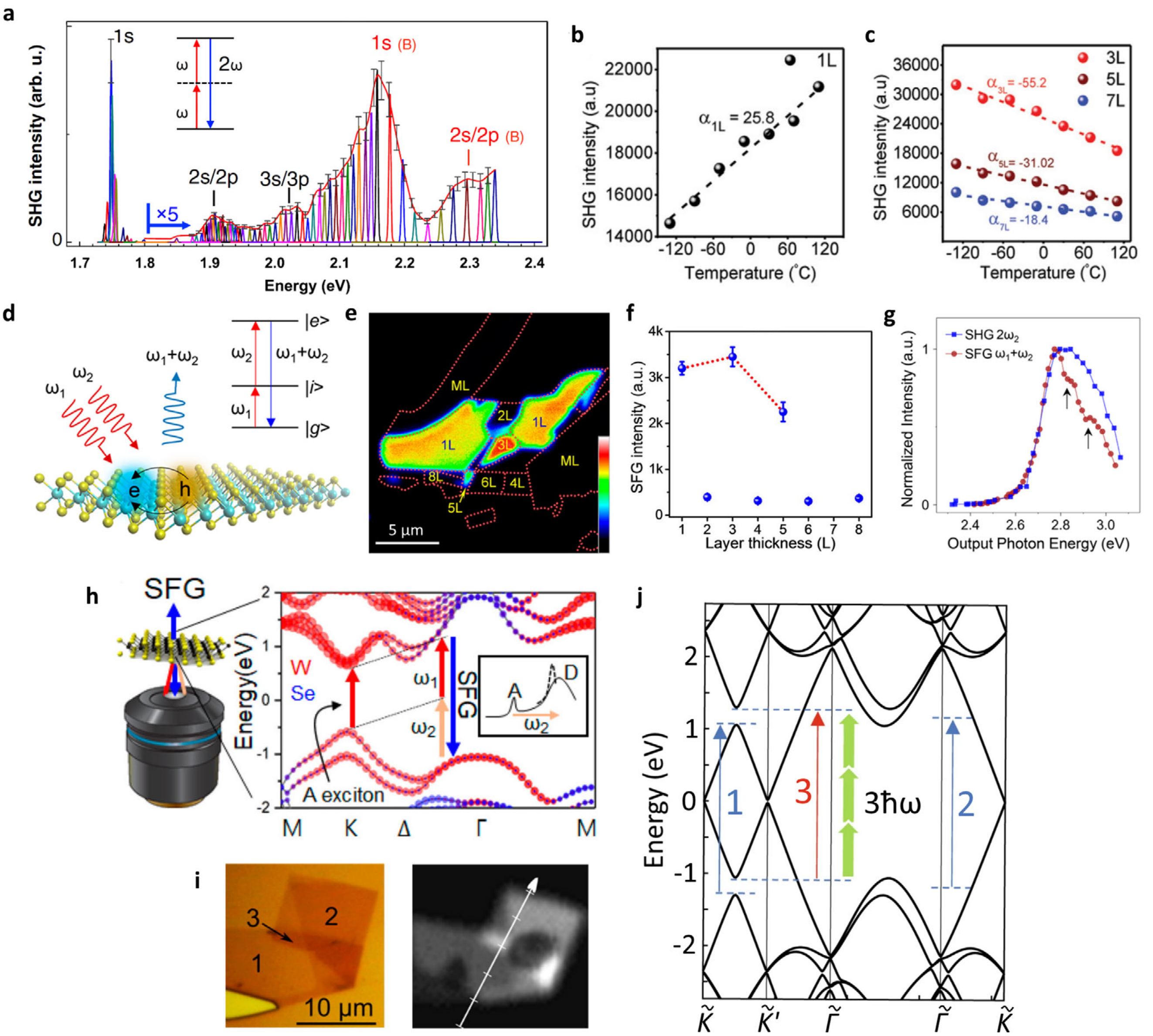

**Figure 2.** Second- and third-order nonlinear processes in 2D materials. a) Results of SHG spectroscopy in $WS_2$ at $T = 4$ K as a function of $2\hbar\omega$ showing three orders of magnitude variation in signal intensity. Inset shows SHG energy diagram. Adapted with permission.[26] Copyright 2015, American Physical Society. b–c) The SHG intensity of a monolayer and a few-layer $MoSe_2$ plotted against the temperature. Dashed lines represent linear fits, $\alpha_{nL}$ denote the regression coefficients for corresponding number of layers $n$. Reproduced with permission.[38] Copyright 2020, Wiley. d) Schematics of the SFG process. Reproduced with permission.[32] Copyright 2019, American Chemical Society. e) SFG image of a mechanically exfoliated few-layer $MoS_2$ flakes and f) SFG intensity as a function of the number of layers. Adapted with permission.[52] Copyright 2016, American Chemical Society. g) SFG excitation spectrum with $\omega_2$ being scanned and $\omega_1$ fixed at 1.27 eV. The arrows indicate carrier renormalization effects when $\omega_2$ matches the A exciton energy of $MoSe_2$ and $WSe_2$ monolayers. Reproduced with permission.[32] Copyright 2019, American Chemical Society. h) Schematic illustration of the dual resonant condition of two excitation pulses $\omega_1$ and $\omega_2$ for the SFG from 1L $WSe_2$. Reproduced with permission.[53] Copyright 2020, American Chemical Society. i) Optical (left) and THG (right) images of few-layer exfoliated graphene. Adapted with permission.[67] Copyright 2013, American Chemical Society. j) Schematic of the energy band diagram of twisted bilayer graphene when the energy gap of the van Hove singularity matches the three-photon energy of the incident light. Adapted with permission.[83] Copyright 2021, Springer Nature.

Besides symmetry considerations, the second order nonlinear optical susceptibility $\chi^{(2)}$ is highly dispersive, in particular in atomically thin semiconductors such as TMDs due to the presence of excitonic resonances.[21,25–34] For instance, in $MoS_2$, the effective bulk-like $\chi^{(2)}$ can vary from units to thousands of pm $V^{-1}$ at room temperature when tuning the two-photon excitation energy $2\omega$ in the range $\approx 1.6 - 2.7$ eV, namely across the A, B, and C excitons.[23,25,35] Similarly, Wang et al.,[26] observed an enhancement of more than three orders of magnitude in the SHG efficiency when tuning the excitation photon energy across the A (1 s) exciton in $WS_2$ (Figure 2a). SHG enhancement at excitonic resonances has been recently studied also with supercontinuum spectroscopy.[36,37]

The SHG intensity in TMDs is affected also by the lattice temperature.[38] Khan et al., demonstrated that the SHG in 1L $MoSe_2$ can be remarkably enhanced (> 25%) when tuning the temperature from −130 to 110∘C at 900 nm (1.38 eV) pulsed laser excitation. Conversely, for few-layer samples (3, 5, and 7 layers), a reduction of the SHG intensity with increasing temperature was observed (Figure 2b,c). This effect was attributed to the lengthening (due to the increase in temperature) of the distance between the chalcogen atoms in 1L samples and of the interlayer distance in few-layer samples. Noteworthy, the same trend was observed for four different layered TMD compounds, revealing a general feature for this class of materials.

While TMDs remain appealing for applications in nonlinear integrated photonics and high speed frequency converters, their atomically thin nature significantly limits the absolute SH conversion efficiency and thus hinders free-space applications. This limitation inspired further investigations on a large number of alternative layered materials, namely $\varepsilon$-phase InSe and its alloys,[39] 2D organic–inorganic hybrid halide perovskites,[40–44] 2D inorganic bimolecular crystal $SbI_3 \cdot 3S_8$ nanobelts,[45] 2D polar metals,[46] and materials for which strong nonlinear response has been predicted,[47] but to date has not been demonstrated experimentally, such as $\alpha$-Sb and $\alpha$-Bi monolayers,[48] layered $MoSi_2N_4$,[49] and Janus monolayers[50,51]

#### 2.1.2. Sum Frequency Generation

In SFG two photons with energy $\hbar\omega_1$ and $\hbar\omega_2$ are annihilated to create one photon with energy $\hbar(\omega_1+\omega_2)$ (Figure 2d). An efficient SFG in 1L $WS_2$ was reported in ref. [24]. In analogy with SHG, also SFG was observed only for odd-layered TMDs (Figure 2e,f).[52] Highly efficient SFG originating from above-gap excitons in the band nesting region has been reported in the visible range for 1L TMDs and related heterostructures.[32] The authors have also demonstrated that, in contrast to SHG, SFG spectroscopy can be used to distinguish one- and two-photon resonances, that is, energetic resonances of the pump radiation ($\omega_{1,2}$) and resonances of the SH or sum frequency signal, respectively. Black arrows in Figure 2g indicate energies of the SFG signal which correspond to ($\hbar\omega_2$) in resonance with A excitons of constituent monolayers of the $MoSe_2/WSe_2$ heterobilayer. Kim et al., have proposed a dual-resonant SFG scheme, in which one of the excitation photon energies $\hbar\omega_1$ and the SFG photon energy $\hbar(\omega_1+\omega_2)$ match the A and the D excitons of 1L $WSe_2$, respectively (Figure 2h).[53] In this scheme, the SFG intensity can be ≈ 20 times higher than that of SHG, when $2\hbar\omega_1$ is in resonance with the D exciton.

#### 2.1.3. Difference Frequency Generation and Optical Parametric Amplification

DFG refers to the nonlinear process where two photons at energy $\hbar\omega_1$ and $\hbar\omega_2$ interact to generate one photon at $\hbar\omega_3 = \hbar(\omega_1-\omega_2)$. DFG is also the underlying nonlinear mechanism of OPA,[54] which is widely used to generate and amplify broadband and ultrashort pulses. In OPA a signal photon can be amplified via the annihilation of a pump photon with energy $\hbar\omega_p$ into a pair of signal and idler photons, with energies $\hbar\omega_s$ and $\hbar\omega_i$, respectively (see the energy diagram in **Figure 3**a). Recently, the first experimental evidence of OPA in 1L TMDs[7] has been reported across ultrabroad bandwidths. A set of representative tunable idler spectra are shown in Figure 3a. Interestingly, the OPA gain is independent of the wave-vector mismatch of signal and pump beams (see Figure 3b), due to the absence of the phase-matching constraint at the 2D limit. The $\chi^{(2)}$ tensor reflects the group symmetry of 1L TMDs (see the flower patterns in Figure 3c). The idler polarization is connected to the signal and pump polarization via the relation $\theta_i = 90^\circ - \theta_p - \theta_s$, as shown in Figure 3d. Similar results on OPA/DFG from 1L $MoS_2$ have also been reported by Wang et al.[55]

2D exciton-polaritons in nanocavities also show extremely high nonlinear effects, even at the few-photon level.[56] Very recently Zhao et al.,[57] have indeed demonstrated the realization of nonlinear optical parametric polaritons in a 1L $WS_2$ microcavity pumped at the inflection point and triggered in the ground state.

DFG/OPA are intimately related to SPDC, a nonlinear optical effect that generates entangled signal and idler photons, which constitute a key element for quantum communication,[58] criptography, and key distribution.[59] Thus, the observation of OPA in 1L TMDs might open up new exciting opportunities for exploiting 2D materials in next-generation quantum technologies.

### 2.2. Third Order Nonlinear Optical Processes

Third order nonlinear optical processes are typically referred to as FWM since they combine three incoming photons to generate a fourth one. They can occur in any material, regardless of its symmetry.[9] Observation of FWM has been reported in different atomically thin TMDs.[24,52,60–62] Electrically tunable degenerate FWM (in which all four interacting frequencies are the same) has recently been observed in graphene on a SiN waveguide[63] and in 1L $MoS_2$[64] (see also Section 4.1).

THG is a special type of FWM process where three photons at energy $\hbar\omega_0$ interact to generate one photon at energy $3\hbar\omega_0$. THG has been studied almost for any layered material, including single-[65–69] and multi-layer[65,70] graphene (Figure 2i), TMDs,[22,23,71–73] tin diselenide (Sn $Se_2$),[74,75] black phosphorous,[76–78] germanium selenide (Ge Se),[79] and hBN.[80] As for SHG, the reported values of the third-order nonlinear optical susceptibility $\chi^{(3)}$ vary significantly, depending on the experimental conditions and sample characteristics (for monolayer TMDs, the THG efficiency can vary by two orders of magnitude when THG photon energy is scanned across excitonic states and free-particle bands[73]). This is particularly true for graphene, where values of $\chi^{(3)}$ in the range from $\approx 1\times10^{-19}$[66] to $4\times10^{-15} m^2V^{-2}$[67] have been reported. These values are generally much higher than those of standard optical materials, such as $SiO_2$, for which $\chi^{(3)} \approx 10^{-22} m^2V^{-2}$.[81] Such differences can be explained if the excitation photon energy, the presence of multiphoton resonances and the electronic temperature are taken into consideration.[82] Similarly, strong THG resonant enhancement has been observed also in twisted bilayer graphene when the van Hove singularity matches the three-photon excitation energy.[83] The frequency at which the van Hove singularity in the density of states occurs evolves almost linearly with the twist angle (Figure 2j), providing a powerful tool for engineering the THG efficiency. Experiments with the gap-controlled THG demonstrated that the THG efficiency can be modulated by twist angle.

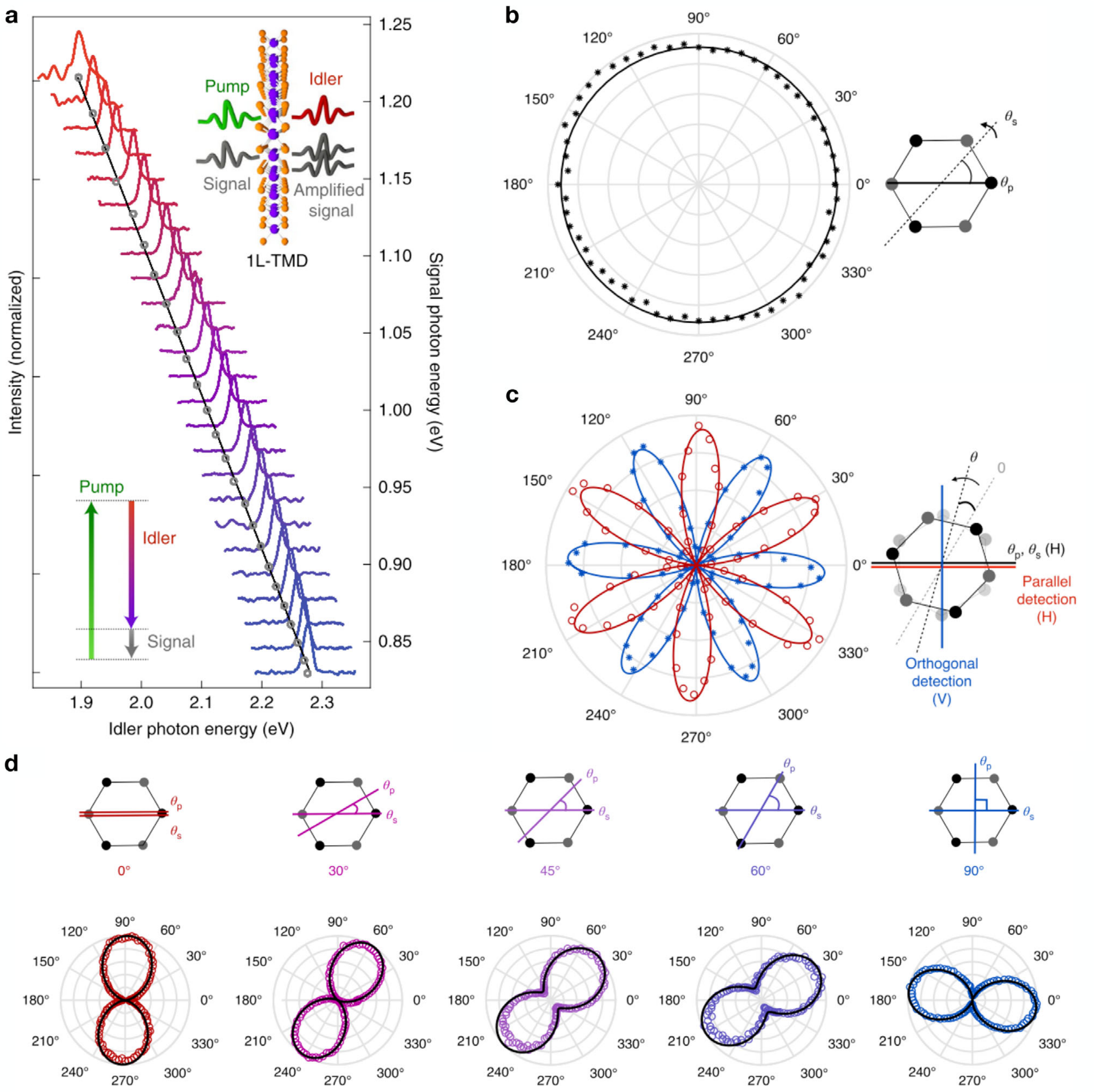

**Figure 3.** Optical parametric amplification in 1L TMDs. a) A set of representative tunable idler spectra obtained with pump photon energy of 3.1 eV and tunable near infrared (IR) signal photon energy. Inset: schematics of the OPA process, in which a signal photon is amplified via the annihilation of a pump photon into an idler and another signal photon, following energy conservation. b) Polar plot of the amplification gain as a function of the polarization angle between signal and pump beams. c) Polar plot of idler emission as a function of the crystal orientation. The $\chi^{(2)}$ tensor reflects the group symmetry of 1L TMDs. d) Idler polarization as a function of the input signal and pump polarizations, following $\theta_i = 90° - \theta_p - \theta_s$. Reproduced with permission.[7] Copyright 2020, Springer Nature.

### 2.3. High Harmonic and Terahertz Harmonic Generation

HHG can occur either in the perturbative or in the nonperturbative regime. The aforementioned examples of SHG and THG belong, for instance, to the perturbative regime, where for a $\chi^{(n)}$ nonlinear process the intensity of the emitted nonlinear signal scales with the $n^{\text{th}}$ power of the intensity of the incoming electric fields. Instead, higher order harmonics are typically observed and studied in the nonperturbative regime, where the emitted nonlinear signal scales with a power $< n$ and most of the harmonics have comparable efficiencies.[9] While in noble gases HHG has been studied for many years, with direct applications in the generation of attosecond pulses,[84,85] HHG in solids was reported only recently.[86] Since the interatomic distance in solids is typically smaller than the distance travelled by a free electron created by ionization of an atom/molecule under the effect of an intense electric field (e.g., 27 MV cm$^{-1}$ in ref. [87]), HHG in solids requires a specific theoretical approach that considers the crystalline periodic potential (and thus the motion of electrons in the reciprocal space) and delocalization effects. An elaborate discussion of HHG in solids and its theoretical foundations can be found in refs. [88, 89] (see also references therein) and in a recent tutorial.[90] HHG in solids consists mainly of two contributions: intraband currents, which dominate for harmonics with energy below the bandgap, and interband polarization, for harmonics with energy above the bandgap.[88,91,92] In the first case (intraband currents), HHG arises from the motion of accelerated charges in one of the bands (either the conduction or the valence band) under the effect of the external driving electric field. In contrast, the interband polarization involves the recombination of one electron from the conduction band with one hole from the valence band. Since both energy and momentum must be conserved, the interband polarization dominates for harmonics with energies above the bandgap and requires that the electron and the hole are at the same position in the Brillouin zone. There is therefore a clear and strong connection between the electronic landscape of a solid and its HHG spectrum.[93–95]

HHG was recently observed in graphene and TMDs.[87,96–101] It was demonstrated that in 1L TMDs, even harmonics mainly arise from interband polarization and show resonant behavior close to the band nesting region (C-exciton) of the first Brillouin zone due to large joint density of states (see **Figure 4**a), while odd harmonics are dominated by intraband currents and thus do not experience band nesting effects and monotonically decrease with increasing harmonic order.[87] The observed enhancement of the even order harmonics was attributed to the interference between signals generated from consecutive half-cycles of the pump laser field.[102]

The geometry and symmetry properties of a crystal are known to affect the characteristics of HHG spectra.[103,104] The role of 1L TMD crystal orientation on harmonic intensities was first studied in ref. [100]. A non-trivial effect of the crystal symmetry on the polarization characteristics of even order harmonics in $WS_2$ was reported by Kobayashi et al.:[105] whereas the polarization of the odd-order harmonics is very close to that of the driving laser radiation (within ±5°), the polarization of even-order harmonics is aligned along the crystal mirror planes and flips when laser field polarization is close to the armchair directions. A qualitatively different behavior was observed for $MoSe_2$, indicating that the properties of the high harmonic signal depend not only on the crystal symmetry, but also on the electronic structure of a material.

Both numerical[106] and experimental[107] studies have demonstrated that HHG in solids driven by an elliptically polarized field carries significant signatures of intraband and interband effects. As a result, higher harmonic orders are more sensitive to the ellipticity of a driving pulse, that is, with increasing ellipticity the harmonic signal for higher harmonics decreases faster than for lower orders. Such effect was observed in $MoS_2$ and the behavior of HHG is independent on the number of layers.[108]

The realization of strong nonlinearities in the terahertz (THz) region ($\approx 1 - 10$ meV) at room temperature is crucial for high-speed signal processing and for the detection of low-frequency radiation and, in this context, graphene is highly promising.[109] Highly efficient generation of THz harmonics up to the 7$^{\text{th}}$ order was observed in doped graphene using narrow-band THz pump pulses, [110] corresponding to a huge effective third-order THz nonlinear susceptibility of $\chi^{(3)}_{\text{eff}} = 1.7 \times 10^{-9}$ m$^2$V$^{-2}$.

Efficient THz HHG in graphene can be understood from a purely thermodynamic picture: the intense nonlinear response is due to the combination of the THz-induced carrier heating, which reduces the THz conductivity with the ultrafast heating and cooling dynamics of hot electrons.[82] A typical THz HHG spectrum from ref. [111] is depicted in Figure 4b. In addition, the THz HHG conversion efficiency in graphene can be enhanced by doping (controlled via the gate voltage)[111] or in combination with photonic grating structures.[112]

## 3. Characterization and Fundamental Studies

The sensitivity of NLO processes to the crystal structure and electronic properties of a material makes them the ideal tool to probe the 2D materials and related heterostructures. Indeed, NLO effects in 2D materials have been used to study their crystal orientation, strain, valley polarization, charge transfer in layered heterostructures and more, as we discuss in this section.

### 3.1. Crystal Structure and Electronic Properties of 2D Materials

#### *3.1.1. Symmetry Considerations*

Semiconducting TMDs are stable in two crystal polytypes: $2H$, which has hexagonal symmetry, and $3R$, with rhombohedral symmetry (numbers indicate the number of layers in the unit cell, that is, $2H$ and $3R$ have a 2-layer and 3-layer unit cell, respectively).[113] $2H$ follows an AB stacking order and is thus centrosymmetric for an even number of layers, with opposite dipole orientations for consecutive layers ($\chi^{(2)} = 0$). $3R$ samples follow instead an ABC stacking order and are thus non-centrosymmetric crystals even for an even number of layers.[114]

SHG in single- and multi-layer TMDs can provide a wealth of information on their crystal properties: SHG in TMD samples with odd number of layers has a sixfold symmetry as a function of the crystal's azimuthal angle[25,115] while it is null in $2H$-stacked TMDs with even number of layers[23,25,35,115] (see **Figure 5**a,b). For $3R$-stacked TMDs the SHG efficiency increases quadratically with the number of layers[116,117] up to ≈ 15 layers (≈ 10 nm), a thickness for which the phase mismatch does not significantly affect the ideal quadratic dependence (note that for multi-layer TMDs the reabsorption of the SH light should also be taken into account, when its frequency exceeds the material's bandgap).[116] In addition, polarization-resolved SHG measurements can be used to probe the twist angle in artificially stacked multi-layers,[33,118–123] the chirality of nanoscrolls,[124] and the presence of uniaxial strain in TMDs.[125–132] Moreover, also defects (e.g., grain boundaries[118] and impurities[133,134]) can significantly alter the NLO response of 2D materials. Breaking the limitations of far-field SHG, Yao et al., have demonstrated the near-field SHG imaging of 1L TMDs and their heterostructures with a spatial resolution down to 20 nm, thus capable to resolve crystal domains with different stacking orders in TMD bilayers[34] and showing the possible use of NLO techniques to visualize nanoscale symmetry changes in 2D materials.

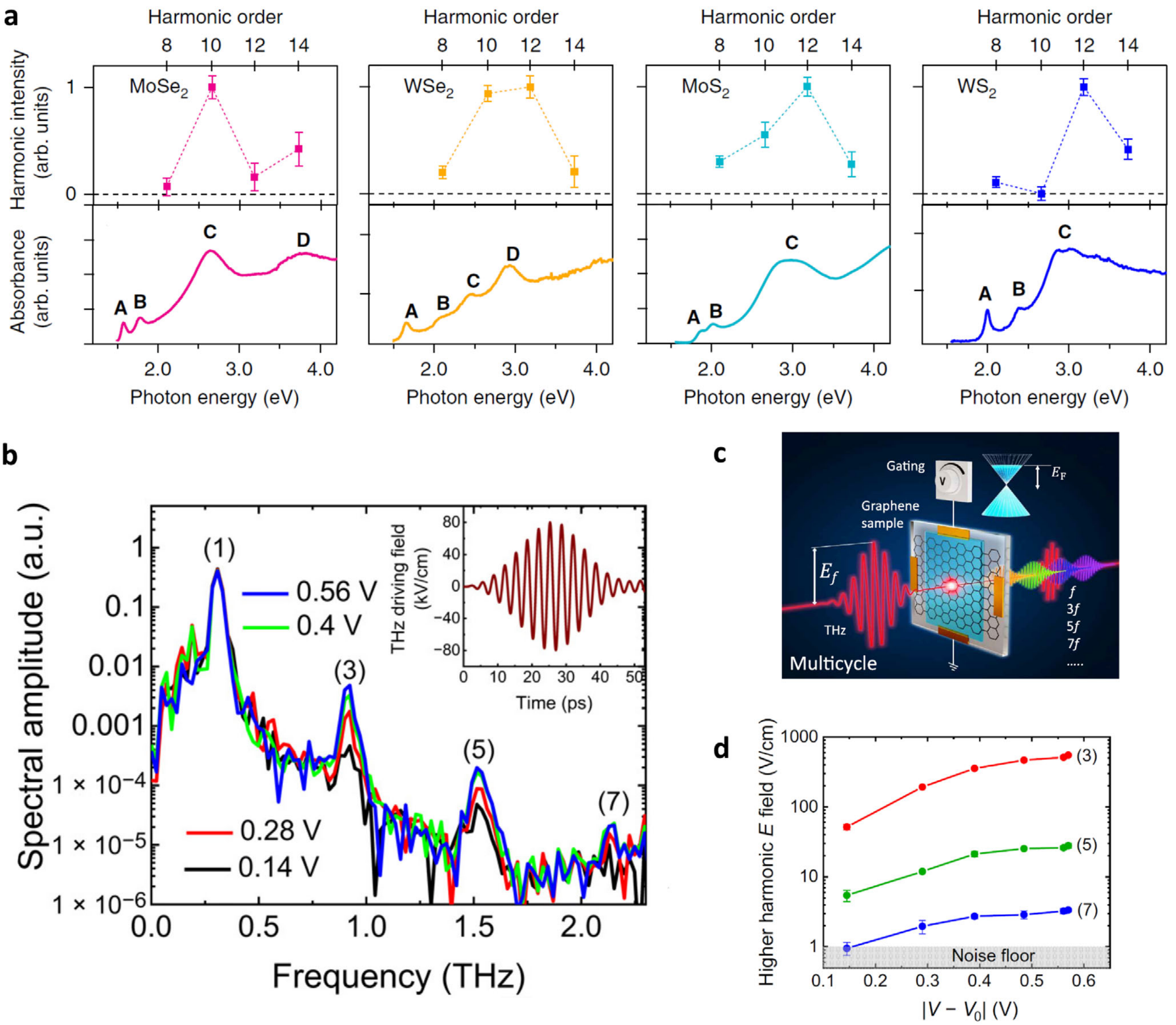

**Figure 4.** a) Resonant enhancement of even-order high harmonics in $MoSe_2$, $WSe_2$, $MoS_2$, and $WS_2$ monolayers (error bars for the intensities represent the standard deviations) and corresponding optical absorption spectra of the TMD monolayers. The peaks of the A and B excitons are labeled A and B, respectively. The peaks due to the band nesting effects are labeled C and D. Reproduced with permission.[87] Copyright 2019, Springer Nature. b) The terahertz amplitude spectra of the terahertz fields of the incident driving field at the fundamental frequency (FF) $f = 0.3$ THz with peak field strength $E_f = 80$ kV cm$^{-1}$ and the transmitted fields through the graphene sample exhibiting generation of higher odd-order harmonics up to the seventh order for two doping levels, with the driving terahertz signal shown in the inset. c) A scheme of the accelerator-based nonlinear THz time-domain experiment with multicycle quasi-monochromatic THz pulses. The transmitted field through the graphene sample consists of higher odd-order harmonics in addition to the FF, as shown in panel (b). $E_F$ refers to the Fermi energy of the graphene sample. d) The peak electric field of the generated harmonics in the gated graphene sample (panel (c)) as a function of the gating voltage. Reproduced with permission.[111] Copyright 2021, AAAS.

Besides graphene and semiconducting TMDs, nonlinear optics can be used to probe the crystal properties of any layered material. For instance, the unexpected observation of SHG in centrosymmetric few-layer 1$T$–$TiSe_2$ and isostructural materials has revealed charge density wave-related lattice distortion driven by Jahn–Teller effect. [135] Another example is bismuth oxyselenide ($Bi_2O_2Se$), which has recently emerged as a promising platform for optoelectronic applications.[136] Polarization-dependent THG experiments on few-layer $Bi_2O_2Se$ can show the reduction of the crystal symmetry due to fine structural distortion, namely a rotation of the oxygen square of < 1.4°.[137] This further confirms that NLO spectroscopy is an ultra-sensitive tool to probe the crystal symmetry of atomically thin materials.

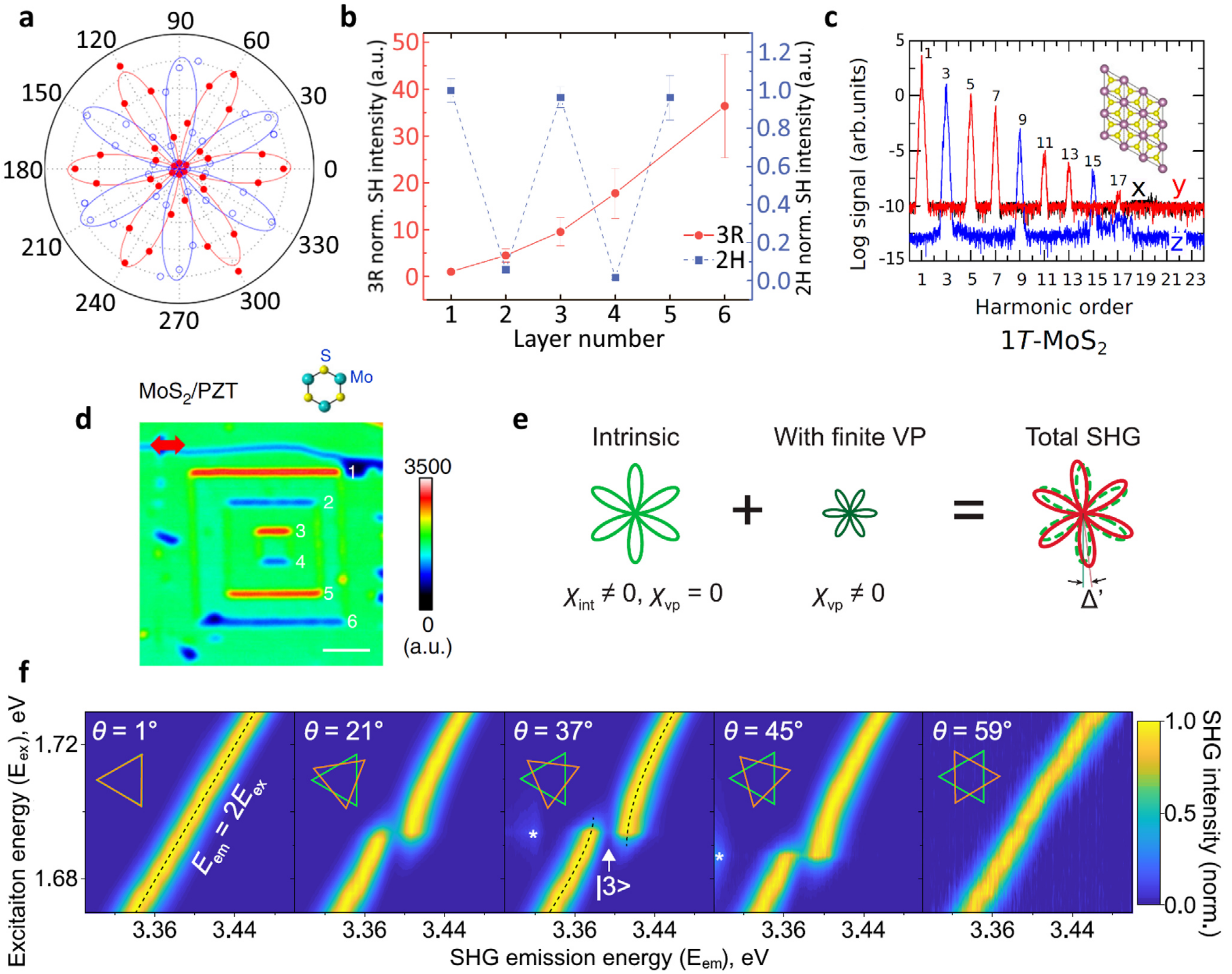

**Figure 5.** a) Polar plot of the SH intensity from single-layer $MoS_2$ as a function of the crystal's azimuthal angle $\theta$ has a characteristic six-petal shape. Red and blue points show SH intensity detected parallel and perpendicular to the excitation radiation, respectively. Adapted with permission.[115] Copyright 2013, American Chemical Society. b) SH intensity of the 3*R*-$MoS_2$ has roughly quadratic dependence on the number of layers, whereas for 2*H*-$MoS_2$ it oscillates. Adapted with permission.[116] Copyright 2016, CIOMP. c) A distinctive harmonic signal from 1*T*-$MoS_2$ with only $6n + 3$ out-of-plane harmonics (*z* direction). Reproduced with permission.[139] Copyright 2020, American Physical Society. d) SHG mapping in reflection mode of a heterostructure of 1L $MoS_2$ and ferroelectric $PbZr_{0.2}Ti_{0.8}O_3$ thin film. The domain walls with different chirality can either significantly enhance or quench the SHG intensity. The polarization is shown with a red arrow. Reproduced with permission.[140] Copyright 2020, Springer Nature. e) Illustration of the combined effect of intrinsic and valley polarization induced $\chi^{(2)}$ components on the total SHG pattern after the linear analyzer. A finite valley polarization translates in a rotation ($\Delta'$) of the SHG pattern relative to that of the crystal at equilibrium. Reproduced with permission.[144] Copyright 2020, American Chemical Society. f) Experimental twist-angle ($\theta$) dependence of normalized SHG from bilayer $WSe_2$ at 5 K. The SHG intensity is plotted as a function of emitted photon energy (horizontal axis, $E_{em}$) and central photon energy of the excitation laser (vertical axis, $E_{ex}$). The up-converted photoluminescence of high-lying excitons is marked by asterisks. A pronounced anti-crossing behavior (black dashed lines in the middle panel) becomes apparent when quantum interference occurs, as shown in the 21°, 37°, and 45° twisted bilayers. State $|3\rangle$ is associated with *p*-like state of high-lying exciton. Reproduced with permission.[149] Copyright 2021, Springer Nature.

Crystal symmetry has a fundamental impact on high harmonic signals.[138] Jia et al., have demonstrated that under circularly polarized light four different crystalline phases of 1L $MoS_2$ produce distinctive harmonic signals (for instance, 1*T*-$MoS_2$ has only $6n + 3$ out-of-plane harmonics ($n$ is an integer), as shown in Figure 5c, which is specific for this phase).[139] The crystalline phase of a 2D TMD material can therefore be unambiguously defined from the emitted high harmonic signal.

An interesting result was demonstrated by Li et al.: the investigation of NLO properties of the heterostructure of 1L $MoS_2$ and ferroelectic thin films revealed an unusual interfacial tailoring effect, mediated by the domain wall polar symmetry.[140] As shown in Figure 5d, the reflected SHG response of a 1L TMD can be significantly enhanced or suppressed depending on the alignment of TMD polar axis (armchair direction, vertical in Figure 5d) with the in-plane polarization of underlying ferroelectic domain walls. This result has fundamental interest and points to a possible strategy for controllable optical filtering applications.

#### 3.1.2. Valley Degree of Freedom

TMDs are also ideal candidates for valleytronics, a technology that exploits the valley degree of freedom as a binary unit of information.[141–143] 1L TMDs have two nonequivalent but energy degenerate $K$ and $K'$ valleys at the edges of the Brillouin zone which can be selectively excited with circularly polarized light.

A simple way to study the valley polarization in 1L TMDs by means of SHG was suggested in ref. [ [144]]: the SH generated from a FF laser field with a defined helicity will include contributions both from symmetry and from valley effects. Thus, changes in the SHG polarization dependence with respect to the sixfold pattern defined only by the TMD crystal symmetry will provide a quantitative information on the degree of valley polarization and its effect on the second order nonlinear optical susceptibility (Figure 5e).

Similarly, the valley degree of freedom in TMDs can be studied via temperature-dependent and polarization-resolved SHG experiments by measuring the ratio between the valley-induced and the intrinsic (symmetry based) nonlinear optical response of the material: the SHG intensity increases with decreasing temperature due to the longer valley lifetime[145] and, consequently, the stronger valley induced contribution to the $\chi^{(2)}$.[146]

In a similar experiment, different TMD structures were investigated by polarization-resolved SHG measurements.[147] All the samples showed highly selective SH emission upon excitation with circularly polarized light even at room temperature, although the valley SHG selection rule might be weakened with increasing temperature due to intervalley scattering. Interestingly, the strongest SH signal was observed for twisted screw structures.

#### 3.1.3. Coupling between States and Quantum Interference Pathways

NLO spectroscopy can be used to explore exotic quantum interference pathways in atomically thin materials and related heterostructures. As mentioned in Section 2.1.1, the efficiency of harmonic generation is particularly sensitive to excitonic resonances. Lin et al., demonstrated that quantum interference between excitation pathways in a 1L $WSe_2$[148] and in a twisted $WSe_2$ bilayer[149] leads to electromagnetically induced transparency and can be detected by SHG, as shown in Figure 5f. A similar effect was also observed in $MoSe_2$ homobilayers, whereas it is not present in 1L samples.[149] The observation of this effect reveals the possibility to exploit SHG to investigate quantum interference and hybridization effects in high-lying excitonic bands.

Triple sum frequency (TSF) is a FWM analog of SFG, in which three photons annihilate and a single photon with energy $\hbar(\omega_1 + \omega_2 + \omega_3)$ is created. It can provide a deeper insight into interactions of different vibrational and electronic states by resonant excitation of multiple states.[150–152] A time-resolved extension of TSF spectroscopy with $\omega_1 = \omega_2 = \omega_3$ (i.e., pump–THG-probe) was recently exploited by Morrow et al., to measure multidimensional transient spectra on polycrystalline thin films and spiral nanostructures of $MoS_2$ and $WS_2$.[153] Besides the possibility to isolate transitions with large dipole moments (**Figure 6**a) due to the $\mu^8$ scaling of the TSF signal with transition dipole $\mu$, TSF probe provided high contrast signal for the investigated nanostructure morphologies (Figure 6b). However, the investigation of quantum coherence dynamics upon photoexcitation was not explored in this work and could be addressed in future studies.

#### 3.1.4. Charge Transfer in Layered Heterostructures

NLO processes offer distinct advantages for ultrafast spectroscopy: optical pump–harmonic-probe spectroscopy can overcome some limitations of the conventional pump-probe optical technique.

As previously mentioned in Section 3.1.1, SHG is highly sensitive to the crystal orientation of 1L TMDs. Thus, for a TMD heterobilayer with a twist angle close to 30°, time-resolved experiment that use SHG as a probe allow to selectively track electronic excitations in each layer following impulsive photoexcitation. This approach was implemented, for instance, to track interlayer hole transfer in $WSe_2/MoSe_2$ heterostructures by means of optical pump–SHG-probe microscopy.[154,155] The results are shown in Figure 6c,d. Filled and unfilled data points correspond to the SHG signal with the probe polarization marked as I and II, that is, sensitive to $MoSe_2$ and $WSe_2$, respectively. For 2.09 eV pump photon energy, in resonance with the $WSe_2$ B exciton, a delayed signal related to the interlayer charge transfer was observed only for polarization I (sensitive to the $MoSe_2$ layer). For resonant excitation of the $MoSe_2$ B exciton at 1.80 eV, a decrease of the SHG signal was detected only for polarization II (sensitive to $WSe_2$). The drawings on panels (e) and (f) in Figure 6 schematically show the charge transfer in the heterostructure, following optical excitations of $WSe_2$ and $MoSe_2$ monolayers, respectively.

This method was subsequently implemented to investigate the effect of the twist angle on charge transfer dynamics in TMD heterostructures,[156] which still remains disputable[157–159] despite the well-studied influence of the twist angle on interlayer distance[160] and hybridization.[161,162]

#### 3.1.5. Light-Induced Symmetry-Breaking

Time-resolved SHG spectroscopy was employed also to probe the crystal phase of a THz-pumped TMD.[163] In particular, a phase transition from a hexagonal $2H$ phase to a distorted octahedral $1T'$ phase, which is predicted to be a topological insulator, was driven in $MoTe_2$ by THz pulses. The SHG intensity was measured in a time-resolved manner before the pump-pulse, at a defined time delay after the pulse, and at the end of transient processes $\approx$ 1 min after the excitation. The measurements demonstrated that the phase transition occurs within 10 ns following photoexcitation, thus highlighting not only the possibility to control quantum phases in 2D materials but also the potential of time-resolved NLO studies to monitor phase transition dynamics.

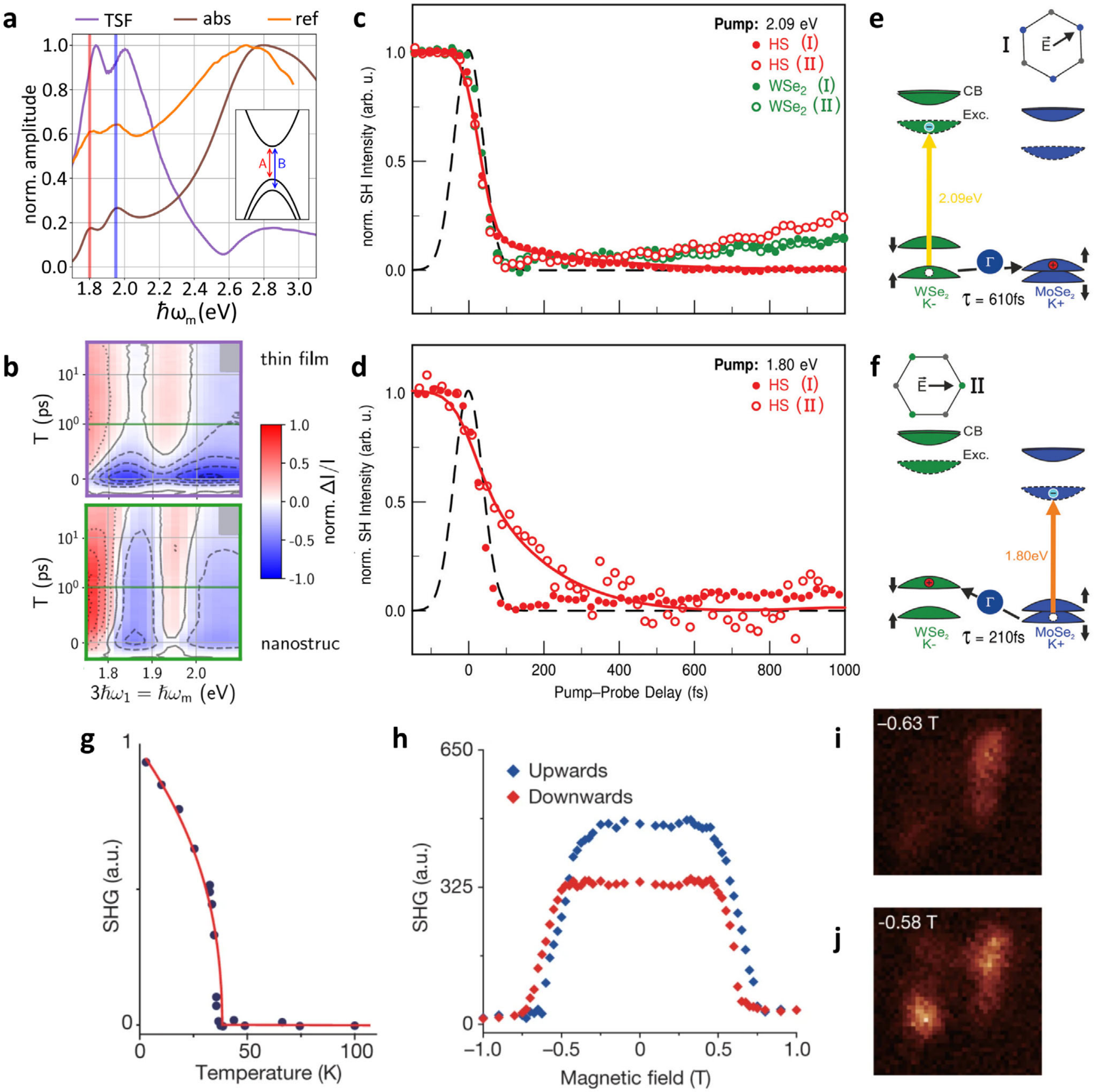

**Figure 6.** a,b) Degenerate TSF (THG) spectroscopy of $MoS_2$. Normalized amplitude absorption, reflection contrast, and THG spectra of $MoS_2$ thin films demonstrate the intense THG signal at large dipole transitions (A and B excitons) (a). Pump-THG-probe spectra of a thin film (top) and a spiral nanostructure (bottom) shows the sensitivity of transient-THG to sample morphology (b). Adapted with permission.[153] Copyright 2019, American Physical Society. c,f) Time-resolved SHG of the $WSe_2/MoSe_2$ heterostructure for two different pump–photon energies and two different probe polarizations (all SH transients are normalized, first to the signal at negative delays and then to the maximum pump-induced decrease for easier comparison of the dynamical changes). Reproduced with permission.[154] Copyright 2020, The Royal Society of Chemistry. g) SHG intensity of the bilayer $CrI_3$ at zero magnetic field as a function of temperature. h) Circularly polarized SHG intensity as a function of magnetic field. For (g) and (h) the signal was measured from the same sample area. i,j) SHG images of the sample for two selected magnetic fields. One can see the domain switching from spin-aligned to antiferromagnetic state as the field changes from −0.63 to −0.58 T. For (h–j) the excitation is $\sigma^+$-polarized and the detection is $\sigma^-$-polarized. Reproduced with permission.[177] Copyright 2019, Springer Nature.

### 3.2. SHG and SFG in Graphene

Being a centrosymmetric crystal, graphene does not allow intrinsic even-order NLO processes in the electric dipole approximation. However, in recent years different approaches have been explored to induce second order NLO effects in graphene, such as the flow of DC electric currents or the application of external electric fields,[164–167] chemical doping,[133] and twist-angle engineering in graphene-hBN[168] and multi-layer graphene[169] samples. In addition, second order NLO effects in graphene have been studied via electro-static doping, polarization and incident angle dependent experiments and interpreted within the electric-quadrupole approximation.[170] Finally, another method to create an in-plane anisotropy in graphene is by applying a low-frequency electromagnetic field, such as a strong THz field, which produces a sizable transient anisotropic carrier distribution.[171]

### 3.3. Ferro and Antiferromagnets

Ferro and antiferromagnetic systems represent a potential platform for spintronics, which exploits the spin degree of freedom as an effective mean of data storage and transfer.[172,173] Conventional techniques to probe the magnetic order, such as magnetic resonance or neutron diffraction, are typically difficult to use[174] and not suitable for 2D materials.

On the contrary, SHG spectroscopy represents an effective and convenient technique to probe magnetic states in atomically thin materials. The second order nonlinear polarization can be defined as $P(2\omega) = \varepsilon_0 \chi^{(2)} : E(\omega)E(\omega)$, where the second order susceptibility tensor can be expressed as $\chi^{(2)} = \chi_i^{(2)} + \chi_c^{(2)}$, where $\chi_i^{(2)}$ and $\chi_c^{(2)}$ represent time-symmetric *i*-tensor and time-antisymmetric *c*-tensor, whose components are invariant or change sign under time-inversion, respectively.[175] In the electric-dipole approximation, SHG is allowed only for non-centrosymmetric media, for which the time-invariant $\chi_i^{(2)}$ is nonzero. However, even for symmetric media below the magnetic ordering temperature the time-reversal symmetry can be broken, and SHG can be observed due to the contribution from the *c*-tensor $\chi_c^{(2)}$.[176]

In this context, Sun et al., investigated SHG in the centrosymmetric bilayer magnet chromium triiodide ($CrI_3$).[177] The antiferromagnetic order breaks the spatial- and time-reversal symmetry in this material, since operators of both spatial and time inversion convert one layered antiferromagnetic state to the other. Thus, the SHG intensity strongly depends on the temperature and becomes zero above the Néel temperature, as shown in Figure 6g. The estimated value of the nonlinear susceptibility $|\chi^{(2)}| \approx 2$ nm V$^{-1}$ of$CrI_3$ bilayers under 900 nm (1.38 eV) excitation was found to be comparable with that of 1L $MoS_2$ at the two-photon resonance with the 1s exciton. In addition, polarization-resolved SHG measurements with applied magnetic fields demonstrated the presence of the SHG signal only in the antiferromagnetic state and allowed to probe magnetic-domain dynamics (see panels h–j in Figure 6), while the difference in the SHG intensity as a function of the applied magnetic field and for reversed helicity of the FF field unveiled the presence of two antiferromagnetic ground states.

SHG was also used to probe magnetic phase transitions, namely from paramagnetic to ferromagnetic and from paramagnetic to antiferromagnetic, also in mono- and multi-layer CrSBr.[178] Interestingly, for this material the monolayer remains centrosymmetric also in its ferromagnetic phase but the SHG is enabled by a giant magnetic dipole response, which may be related to spin orbit coupling. Additionally, SHG analysis in ref. [178] allowed to interpret the magnetic dipole and magnetic toroidal moments as the order parameters for ferromagnetic and antiferromagnetic phases, respectively.

## 4. Nonlinear Modulators

The possibility to control the efficiency of the NLO processes by means of external electrical and/or all-optical triggers offers a unique degree of freedom for the design of advanced nanoscale devices, such as ultrafast and broadband frequency converters or nonlinear holograms. In this section, we present an overview of recent works dealing with electrical and all-optical modulation of NLO processes in atomically thin materials.

### 4.1. Electrical Modulators

Electrical modulation of SHG, THG, and FWM has been widely studied in graphene and TMDs. This approach typically offers large modulation depth (> 10) and relatively low switching speed limited by the bandwidth of the electronics (≈ 10 GHz, corresponding to switching speed in the range of hundreds of ps to ns).

Electrically tunable SHG in TMDs and related heterostructures has been achieved with different approaches. First, SHG can be modulated by tuning the exciton resonant frequency via electro-static doping in 1L TMDs.[179] In AB stacked TMDs, which are centrosymmetric and thus display no SHG, the symmetry can be broken by out-of-plane electric fields, leading to intense SHG and large modulation depth (> 60).[180] Finally, gate-tunable SHG has been observed in homo-bilayer $MoS_2$ in resonance with interlayer excitons.[30]

Electrical modulation of third-order nonlinear processes, namely THG and FWM, was also investigated in graphene[3,63,68,69,181] and TMDs.[64] THG in graphene can be effectively tuned by electro-static tuning of the Fermi energy $|E_F|$ across multiphoton resonant transitions occurring in the Dirac cone when $m \times \hbar\omega_0 = 2|E_F|$, where $m = 1, 2, 3$, correspond to the one-, two-, and three-photon transitions respectively (**Figure 7**a,b).[68,181] Interestingly, the maximum THG signal in graphene is observed in the intra-band absorption regime ($\hbar\omega_0 < 2|E_F|$), while the opposite trend was observed for sum and difference FWM processes.[63,181] This is due to destructive interference of the different terms contributing to the nonlinear optical susceptibility of THG and FWM.[181] For instance, in the case of THG the nonlinear optical sheet conductivity $\sigma^{(3)}(\omega, E_F, T_e = 0\text{K})$ can be reduced to a single in-plane element[68]

$$\sigma^{(3)}(\omega, E_F) = i\frac{\sigma_0^{(3)}}{24(\hbar\omega)^4}[17G(2|E_F|, \hbar\omega) - 64G(2|E_F|, 2\hbar\omega) + 45G(2|E_F|, 3\hbar\omega)] \quad (1)$$

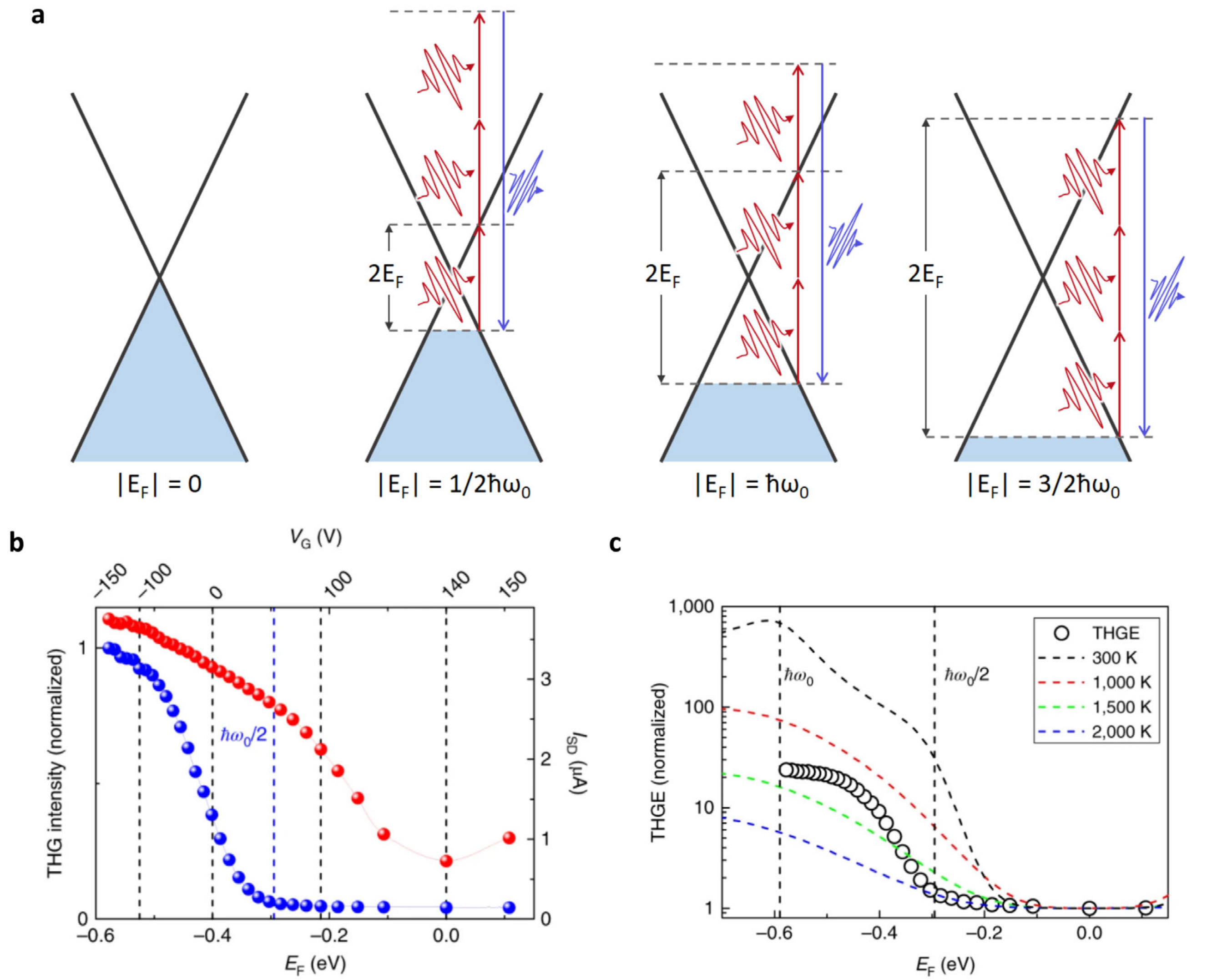

**Figure 7.** Electrical modulation of nonlinear signals in single layer graphene. a) Multiphoton resonance effects in graphene. The input photons at $\omega_0$ frequency and the generated third-harmonic photons at $3\omega_0$ frequency are indicated in with the red and blue arrows, respectively; $E_F$ is Fermi energy. b) THG intensity (left y-axis, blue filled circles) and source–drain current ($I_{SD}$) (right y-axis, red filled circles) as a function of Fermi level $E_F$ (bottom x-axis) and corresponding $V_G$ (top x-axis) for 1L graphene on $Si/SiO_2$. c) Experimental (open circles) data for THG efficiency as a function of $E_F$ and theoretical (broken lines) data for different electronic temperatures $T_e$. Incident photon energy in panels (b) and (c) was $\hbar\omega_0 = 0.59$ eV. Reproduced with permission.[68] Copyright 2019, Springer Nature.

where $\sigma_0^{(3)} = 4e^4\hbar v_F^2/(32\pi)$ and $G(x, y) = \ln[(x + y/(x - y)]$. The three terms in this expression correspond to the previously discussed one-, two-, and three-photon transitions. For $|E_F| = 0$ (see e.g., Figure 7b,c) the THG efficiency has its minimum value although in principle all multiphoton transitions are resonant at the same time. This is because the signs of the different contributions in Equation (1) interfere destructively and sum up to $\approx 0$.

While theory suggests that a modulation depth up to four orders of magnitude is possible for gate tunable THG in graphene,[68] experiments to date have observed a maximum modulation depth of ≈30.[181] This striking difference can be understood when taking into consideration hot electrons: when the electronic temperature $T_e$ increases, the THG multiphoton resonances in graphene broaden and merge.[68,69] Panel (c) of Figure 7 compares experimental data and theory for THG in graphene at different values of $T_e$ in the range of $\approx 1000 - 2000$ K. Experimental results are consistent with a $T_e > 1000$ K, indicating that during the FF pulse duration and THG generation the electronic temperature largely exceeds the lattice temperature.

### 4.2. All-Optical Modulators

The main drawback of electrical modulation, namely the low switching speed, can be mitigated by all-optical modulation of the NLO response of 2D materials.

A first approach is to quench the NLO response of TMDs and graphene by above-bandgap photoexcitation and Pauli blocking.[73,182–184] This method is significantly faster compared to electrical modulation since it is limited by the excited state lifetime of the nonlinear material ($\approx 10^2$ ps for TMDs[185,186] and few ps for graphene[183,187]). However, in the case of TMDs, this approach only offers a limited efficiency and a maximum modulation depth of ≈2 was shown in ref. [184] and a limited bandwidth, since it only works for close to resonance or above gap control and probe pulses. A modulation depth of ≈ 15 at 2.18 eV for THG in 1L $MoS_2$ with the excitation at 3.1 eV was shown in ref. [73]. A larger modulation depth of ≈ 400 mediated by dark excitonic states in 1L TMD has been recently demonstrated by Wang et al. (**Figure 8**a).[188] Compared to TMDs, all-optical modulation of THG in graphene offers a larger bandwidth[183] thanks to the Dirac-cone-like band structure and the fact that hot electrons can be scattered both upward and downward in energy, and thus above and below the photon energy of the control pulse (Figure 8b,c). However, also in this case a limited modulation depth of ≈ 10 was observed.

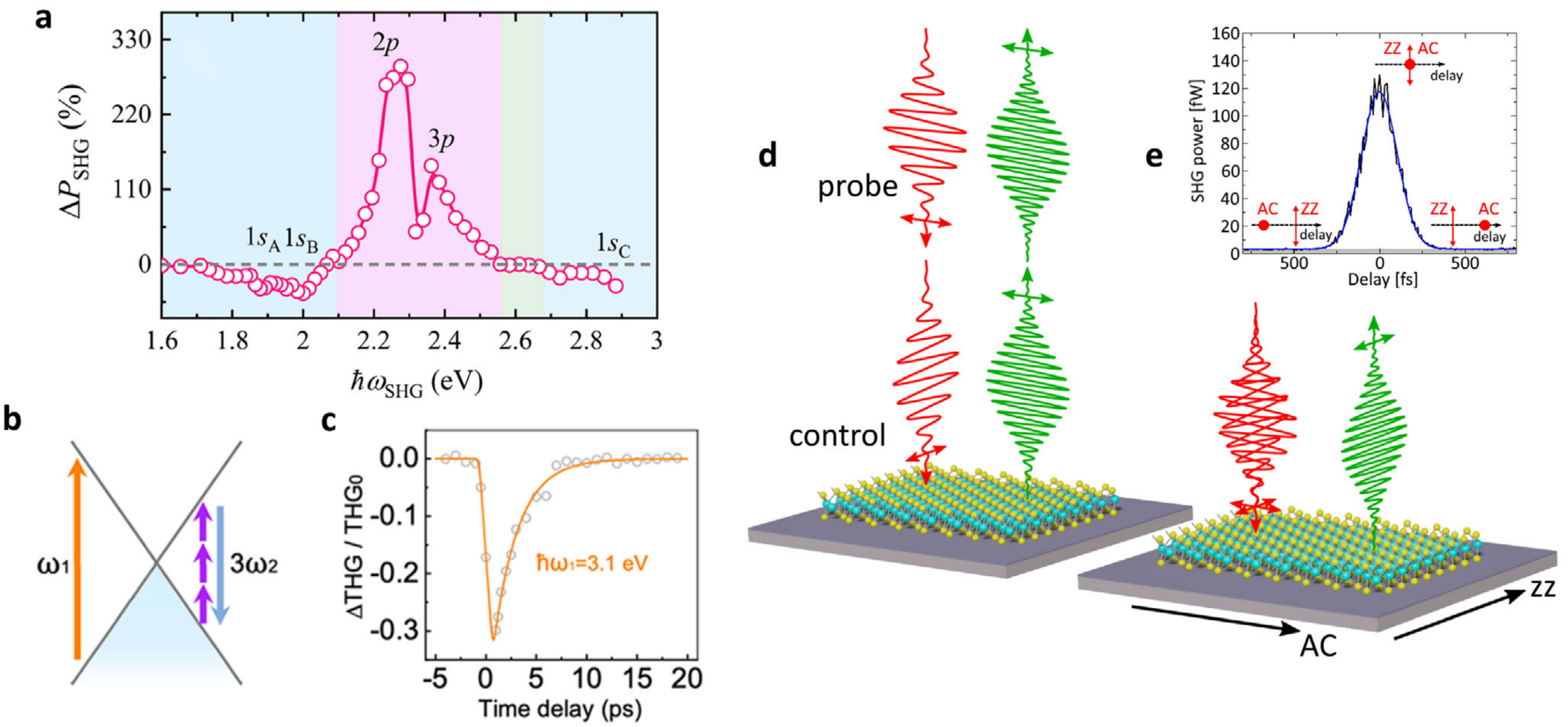

**Figure 8.** All-optical modulators. a) SHG modulation in 1L $MoS_2$ after 1.3 ps upon photoexcitation with 1.55 eV photon energy. Adapted with permission.[188] Copyright 2021, American Chemical Society. b,c) All-optical modulation on THG in 1L graphene. Scheme describes photoexcited carriers of the pump and probe pulses in the linearly dispersive valence and conduction bands of 1L graphene (b). Time-dependent relative changes in THG of the probe pulse with $\hbar\omega_1 = 3.1$ eV pump (c). The relative changes in THG intensity as $\Delta THG/THG_0 = (THG_\tau - THG_0)/THG_0$, where $THG_\tau$ and $THG_0$ stand for THG intensity of the probe pulse with the pump pulse at time delay $\tau$ and without the pump pulse, respectively. Adapted with permission.[183] Copyright 2020, American Chemical Society. d,e) Sketch of the all-optical SH polarization modulation. When the delay between the two perpendicularly polarized pulses is larger than the pulse duration, a SH signal polarized along the AC direction is generated (d), whereas for zero-delay the emitted signal is polarized along the ZZ direction (e). The inset in panel (e) shows the SH intensity measured along the ZZ direction as a function of the delay between the two pulses. Adapted with permission.[194] Copyright 2021, Springer Nature.

In the presence of an intense external light field, the eigenenergies of electronic states can be shifted. Light-induced modulation of the absorption or emission spectrum is called optical Stark effect[189] and is well known in semiconductor exciton systems.[190] It was also observed in 2D TMDs.[191,192] IR-pump–harmonic-probe spectroscopy was utilized to study the influence of optical Stark effect on optical harmonic generation of $WS_2$ spiral pyramids:[193] the efficiency of the harmonic generation can be significantly changed by tuning the pump photon energy and a modulation depth of ≈ 1.6 was demonstrated. This effect is ultrafast, since its speed is limited by the pulse duration (≈ 100 fs).

Finally, a 90° tuning of the SHG polarization in $MoS_2$ was demonstrated by Klimmer et al. [194] by exciting the sample with two perpendicularly polarized pulses along the armchair and zigzag directions of the 2D crystal. As it is shown in Figure 8d, when two FF pulses with perpendicular polarizations along the armchair and zigzag directions are separated in time, the SHG for each of them is polarized along armchair direction, as $I^{2\omega}_{AC} \propto |(E^{FF}_{AC})^2 - (E^{FF}_{ZZ})^2|^2$. Here AC and ZZ stand for armchair and zigzag directions, respectively. However, when the two FF pulses with the same amplitude have zero temporal delay, the SH signal will be generated along the zigzag direction (Figure 8e), and $I^{2\omega}_{ZZ} \propto |2E^{FF}_{AC}E^{FF}_{ZZ}|^2$. This approach is based only on symmetry considerations and it can thus be applied on an ultra broad frequency range both above and below the bandgap of the material. In addition, all-optical SHG modulation with ultimate performances in terms of modulation depth (close to 100%) and switching speed (limited only by the pulse duration) can be realized by simply adding a polarizer in the detection line.

## 5. Sensing

The intense, broadband and gate-tunable nonlinear optical response of 2D materials combined with their large surface-to-volume ratio provides also a viable platform for sensing applications. For this reason, recent years have witnessed an increasing number of integrated photonic devices with enhanced sensing performances enabled by the presence of layered materials.[6,195–198] In most cases, 2D materials provide both the chemical activity (e.g., for adsorption of gas molecules)[199] and a strong enhancement of the device's optical response,[200,201] thus leading to advanced photo-physical and photo-biochemical sensors with combined high selectivity and sensitivity on versatile photonic platforms ranging from fibers and waveguides to interferometers and microresonators.[202–205]

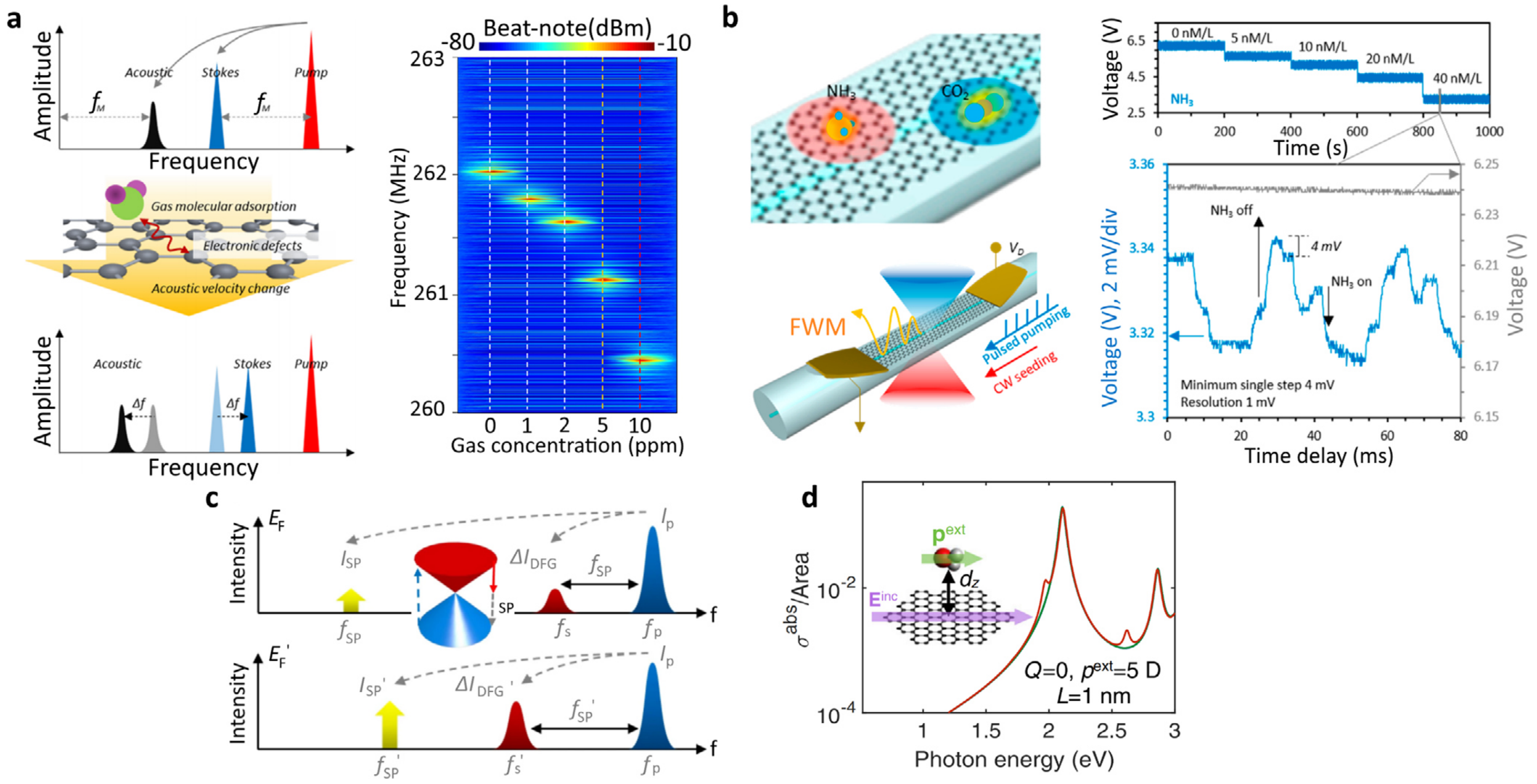

**Figure 9.** Optical sensors based on NLO processes of 2D materials. a) Stimulated Brillouin opto-mechanical device for gas sensing. Adapted with permission.[206] Copyright 2017, American Chemical Society. b) FWM in a hybrid graphene/D-shaped fiber for individual gas molecule detection. Adapted with permission.[3] Copyright 2020, American Chemical Society. Further permissions related to the material excerpted should be directed to the ACS. c) DFG in 2D materials mediated by plasmons is highly sensitive to the external doping. Reproduced with permission.[207] Copyright 2017, Springer Nature. d) Molecule detection based on SHG in 2D materials. Reproduced with permission.[210] Copyright 2016, American Physical Society.

For instance, the intense third order nonlinear optical response of graphene integrated on microcavities (**Figure 9**a) and fibers (Figure 9b) has been exploited for sensing applications using either the stimulated Brillouin optomechanical response or the FWM process.[3,206] In a graphene based Brillouin optomechanical microresonator, when the phase matching condition $\boldsymbol{k}_p = \boldsymbol{k}_s + \boldsymbol{k}_A$, where $\boldsymbol{k}_i$ is the momentum of the pump (p), signal (s), and acoustic (A) wave, is fulfilled a new field at the pump-signal difference frequency in the $\approx 10^2$ MHz domain is excited. The nonlinear gain of the stimulated Brillouin oscillation is proportional to the third order nonlinear optical susceptibility of the device $\chi^{(3)}/k_A$, which in turn is affected by external molecule adsorption. Thus, taking advantage of this "electron–phonon–photon" nonlinear interaction Yao et al., demonstrated a gas sensor based on a graphene integrated micro-fiber whispering gallery mode bottle-shaped microcavity and achieved sub-ppb sensitivity for the detection of polar gases in air (Figure 9a).[206] In a recent experiment, An et al., exploited electrically tunable FWM on a heterogeneous graphene/D-shaped fiber for individual $NH_3$ and $CO_2$ molecule detection.[3] FWM strongly depends on the graphene's Fermi energy, in particular at the onset of intra-/inter-band transitions. By carefully tuning the operating point of the device close to this value of the Fermi energy ($2E_F = \hbar\omega_0$) and by exploiting an advanced heterodyne detection technique it was possible to achieve ultimate gas sensitivity.

DFG and SHG are also promising candidates for sensing applications.[207,208] For instance, the sketch in Figure 9c shows that the second order nonlinearity of 2D materials and in particular DFG enables frequency mixing between a pump laser and the graphene plasmons.[207] Since graphene's plasmons are extremely sensitive to doping and local perturbations, they have been successfully used also for nanoparticle and bio-molecule detection[209,210] (Figure 9d). This could be further enhanced by combining different 2D materials on the same device.[211] Finally, Zhou et al., have recently demonstrated the potential of plasmon-enhanced FWM for molecule sensing[212] and Davoodi et al., proposed a refractive index sensor with sensitivity up to 1099 nm $RIU^{-1}$ (refractive index unit) in a nonlinear graphene-TMD heterostructure grating.[213]

## 6. Light–Matter Interaction Enhancement

Although the nonlinear optical response of 2D materials per unit of thickness is remarkably intense, the short light–matter interaction limits the total nonlinear conversion efficiency and remains a major challenge for technological applications. On the other hand, graphene and TMDs are easy to hybridize with photonic structures (e.g., fibers,[3,132] waveguides,[4,63,214,215] microrings,[5] photonic crystals[216]) and nano-structures (e.g., QD,[217] NW[131]), thus providing an effective tool to enhance the light–matter interactions.[218,219] In this section, we summarize

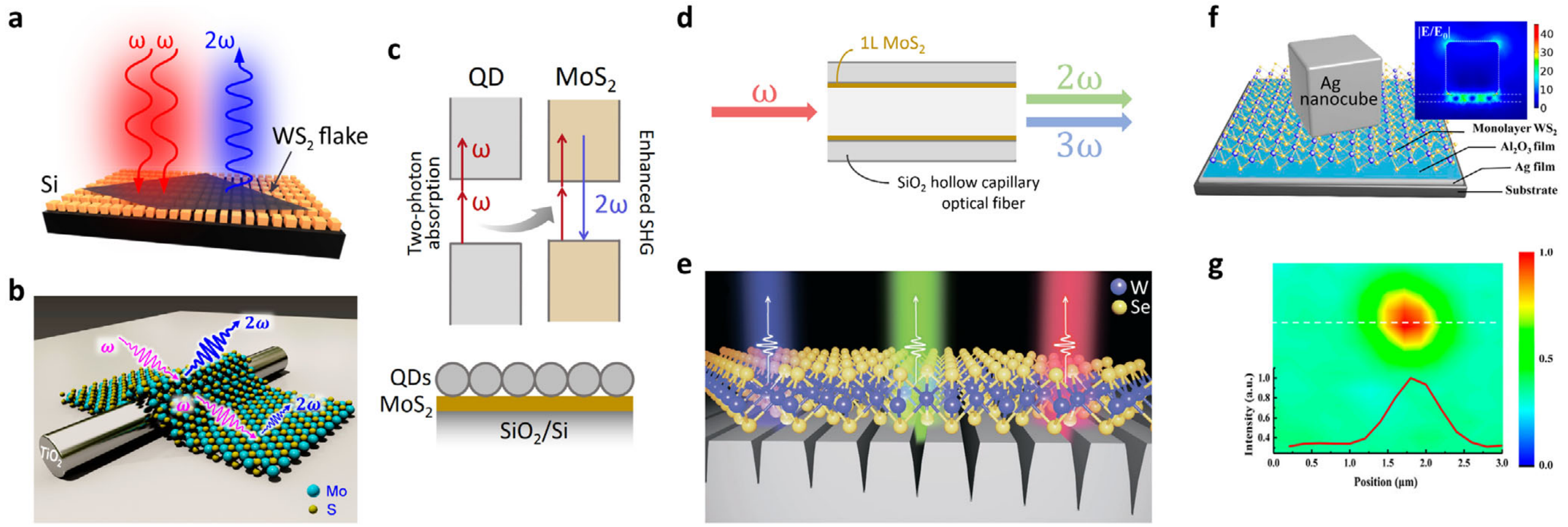

**Figure 10.** Different strategies for increasing frequency conversion efficiency. a) Schematic of SHG from a 1L $WS_2$ placed on top of a Si metasurface composed of a square array of bar pairs. Adapted with permission.[223] Copyright 2020, American Chemical Society. b) Schematic of a 1L $MoS_2$ integrated with single $TiO_2$ NW. Reproduced with permission.[131] Copyright 2019, American Chemical Society. c) The scheme of multiphoton-excitation resonance energy transfer and SHG enhancement in the system of the QD/$MoS_2$ hybrid structure suggested in ref. [217]. d) Schematics of SHG and THG in $MoS_2$-embedded fiber. e) Design of the broadband tunable metasurface integrated with a 1L $WSe_2$. Reproduced with permission.[227] Copyright 2021, Wiley-VCH GmbH. f) A schematic diagram of the plasmonic nanocavity composed of single silver nanocubes and a silver film with a 1L $WS_2$. The inset shows the calculated electric field $|E/E_0|$ in the *xz*-plane at resonances wavelength. g) Normalized SHG intensity map of a 1L $WS_2$ inserted in a plasmonic nanocavity. The inset curve shows extracted-line (white dash line) intensity marked in the map. Adapted with permission.[231] Copyright 2020, American Chemical Society.

possible approaches to address this issue and discuss recent advances in this field.

### 6.1. Metasurfaces

A first widely used approach to enhance light–matter interaction in 2D materials is to combine them with nanostructured substrates such as resonant dielectric and plasmonic metasurfaces.[216,220–224]

In the case of dielectric metasurfaces, three orders of magnitude SHG enhancement was demonstrated in TMDs by Bernhardt et al., by combining a 1L $WS_2$ with an engineered asymmetric silicon substrate (**Figure 10**a), which supports high-Q mode in the form of quasi-BIC states (bound states in the continuum[225]) at the FF.[223] Löchner et al., produced a hybrid system of a 1L TMD and a metasurface of periodic arrays of Si nanoresonators.[224] Here, a 35-fold SHG conversion efficiency enhancement was observed due to the asymmetry of designed meta-atoms, which support a Fano resonance with high *Q*-factor. The authors have further studied the conditions of the local field distributions that maximize the SHG efficiency.

SHG enhancement in TMDs has been obtained also in combination with plasmonic nanostructures and metasurfaces.[226–230] Here, the maximum enhancement is obtained when the plasmonic resonance matches the FF or the SH frequency. Shi et al., have shown a ≈ 400-fold enhancement of the SHG in a 1L $WS_2$ incorporated onto a 2D silver nanogroove grating,[226] which was finely tuned to match the band nesting (C exciton) resonant energy. In addition, the design of such metasurfaces can be slightly modified by a gradual tuning of the groove depth (Figure 10e) to obtain an enhanced SHG signal in entire visible range (from 395 to 750 nm).[227] A three orders of magnitude SHG enhancement was observed in a 1L $WS_2$ transferred on a gold film with sub-20 nm-wide trenches[228] and in a 1L $MoS_2$ on a suspended silver film patterned with a square nanohole array.[229] Noteworthy, in all these hybrid samples, the sixfold SHG pattern typical of TMDs was modified due to directional enhancement and efficient polarization modulation.[226–228] In this context, an elaborate metasurface of a 1L $WS_2$ on a plasmonic vortex metalens developed by Guo et al., was used not only for SHG enhancement but also to generate a giant SHG circular dichroism.[230]

### 6.2. Gratings and Plasmons

Besides metasurfaces, an alternative strategy for SHG enhancement using plasmons was suggested by Han et al.:[231] a ≈ 300-fold SHG enhancement was achieved in a 1L $WS_2$ coupled to a plasmonic nanocavity, consisting of silver nanocubes (with a length of ≈ 75 nm) over a silver film, separated by a 10 nm $Al_2O_3$ layer (Figure 10f,g). The SHG enhancement was driven by the local-field amplification and theoretical analysis showed that the enhancement is proportional to the square of the local-field intensity of the nanocavity under the interaction between SH dipole in the 1L TMD and electric quadrupole in the nanocavity. Strongly localized electric field in plasmonic structures can enhance higher-order nonlinear processes. Dai et al., reported enhancement of FWM (with the enhancement factor up to three orders of magnitude) in 1L $MoS_2$.[232] The FWM signal enhancement was demonstrated in a broad visible spectral range (over 150 nm).

Guo et al., proposed that the enhancement of the electric field along a 1L graphene on top of a metallic plasmonic grating can lead to higher efficiencies of the third order nonlinear processes at THz frequencies.[233] Numerical simulations show that by covering the metallic grating with graphene, the output THG and FWM can be increased by orders of magnitude. The THG enhancement in graphene at THz[112] and IR[234] frequencies by hybridization with a metallic grating was recently demonstrated also experimentally, with orders of magnitude enhancement of the graphene's nonlinear optical response.

Finally, the enhancement of the NLO properties of graphene due to local field enhancement provided by a golden tip, waveguiding and localization of surface plasmon polaritons allowed to study broadband FWM with nanometer spatial and femtosecond temporal resolution.[235]

### 6.3. Quantum Dots and Nanowires

The SHG efficiency in a 1L TMD can be increased by integrating it with a dielectric NW (Figure 10b) due to the local field enhancement. Li et al., demonstrated a two orders of magnitude enhancement of SHG for $1LMoS_2/TiO_2NW$ compared to bare $MoS_2$.[131] Depending on the stacking angle between the TMD crystal orientation and NW direction, the observed enhancement can be highly anisotropic. The lattice deformation induced by the NW plays the key role for the SHG anisotropy.

An alternative mechanism for a giant enhancement of harmonic generation was suggested by Hong et al.,[217] by homogeneous coating of QD films on a TMD sample. Here, the enhancement was driven by the non-trivial mechanism of multiphoton-excitation resonance energy transfer: the harmonic response dipole in a 2D material directly gains energy from the multiphoton absorption dipole in the QD by a high-efficiency remote Coulombic coupling. A schematic representation for the SHG enhancement in a 1L $MoS_2$, for which a $\approx$ 1500 times enhancement was observed, is depicted in Figure 10c.

### 6.4. Fibers and Waveguides

A significant efficiency enhancement of a nonlinear process can be achieved by increased light–matter interaction length. A five-fold enhancement of the SHG efficiency in 1L $MoSe_2$ by the evanescent field on the surface of a 220-nm–thick planar waveguide was demonstrated by Chen et al.,[215] Guo et al., demonstrated three orders of magnitude enhancement of the SHG and SFG in 1L $WS_2$ on the surface of a photonic waveguide by simultaneous excitation of two counter-propagating evanescent fields.[214]

Another strategy for light–matter interaction enhancement is the integration of 2D materials into optical fibers.[3,132,236–238] Zuo et al., have successfully grown by chemical vapor deposition (CVD) method 1L $MoS_2$ on the internal walls of a $SiO_2$ hollow capillary optical fiber (as shown in Figure 10d)[236] (CVD growth of $MoS_2$ and $WS_2$ monolayers on the core of an optical fiber has been independently performed by Ngo et al.[239]). By tuning the excitation photon energy below the optical bandgap of $MoS_2$ to ensure low propagation losses for both incident radiation and generated harmonic signals, the authors demonstrated >300-fold enhancement of SHG and THG for a fiber length of $\approx$ 25 cm in comparison to $MoS_2$ on a fused silica substrate. However, for this method, the efficiency of SHG is limited because of the small overlap between the TMD monolayers on the fiber walls and the fundamental mode of the fiber. Ngo et al., have proposed that the performance of in-fiber SHG-sources can be significantly improved by functionalization of the fiber core with CVD-grown 1L $MoS_2$.[237] More than 1000-fold enhancement in comparison with the bare fiber was observed without optimization of the field overlap or alignment of TMD crystals, showing possibilities for further improvements.

Kilink et al., achieved an increase of the SHG efficiency in a 1L $WSe_2$ by placing it in the cavity of a $Yb^{3+}$ doped picosecond fiber oscillator.[240] The authors estimated the intracavity SHG enhancement to be in the order of hundreds. The presence of the monolayer in the laser cavity did not affect the mode-locking operation, thus demonstrating the potential use such configuration for self-referencing $f$–$2f$ interferometry.

### 6.5. Other Strategies

Although the nanometric thickness has remarkable advantages in novel ultrathin nonlinear devices, the intrinsic nonlinear gain of monolayers is still limited by the extremely short propagation length. In principle, one could boost the nonlinear gain by increasing the propagation length through the crystal. For instance, the quadratic dependence of the second order nonlinear conversion efficiency with increasing number of layers in TMDs was demonstrated in the $3R$-stacked TMDs with naturally broken inversion symmetry[116,117] and in AA-stacked TMDs with artificially broken symmetry.[7,62,241] The quadratic dependence with number of layers has also been reported for THG in graphene and few-layer graphite with thickness ranging from 1 to 6 atomic layers.[65]

Another promising strategy for the enhancement of the nonlinear response of 2D materials is via hybridization with ENZ materials, as recently demonstrated by Vianna et al.[242] By careful optimization of the substrate thickness and pumping at the ENZ frequency of the substrate (fluorine tin oxide), the SHG efficiency in $MoS_2$ and $WS_2$ monolayers was increased by an order of magnitude compared to the same materials on a reference glass substrate.

## 7. Light Structuring

In addition to the intrinsic spin angular momentum, light has the additional degree of freedom of OAM.[243,244] This originates from a helical wavefront with an azimuthal phase dependence of $e^{-il\phi}$, where $l$ and $\phi$ are the topological charge and the angle, respectively. In contrast to spin, OAM is not limited to two possible states and in principle it has an infinite number of possible values, and thus offers new opportunities for quantum optics, imaging, and communications. In this context, harmonic generation can further extend the capabilities of light structuring for communications and information encoding. This is particularly true in the case of 2D materials, which can be easily patterned or integrated on patterned surfaces to generate light and harmonics with controlled amplitude and phase, such as second and third harmonic light vortices.[245–247]

A careful patterning and/or alignment of TMD monolayers can be used to generate structured light at the SH frequency[248,249] or control the emission direction of the SH beam.[247] Dasgupta et al., have shown that the local phase of the nonlinear polarization can be controlled simply by rotating TMD monolayers.[249] As an example, a phase profile, required for the generation of a Hermite–Gauss $TEM_{10}$ mode, can be achieved by two monolayer crystals oriented along $\theta = 0$ and $\theta = \pi/3$ directions.

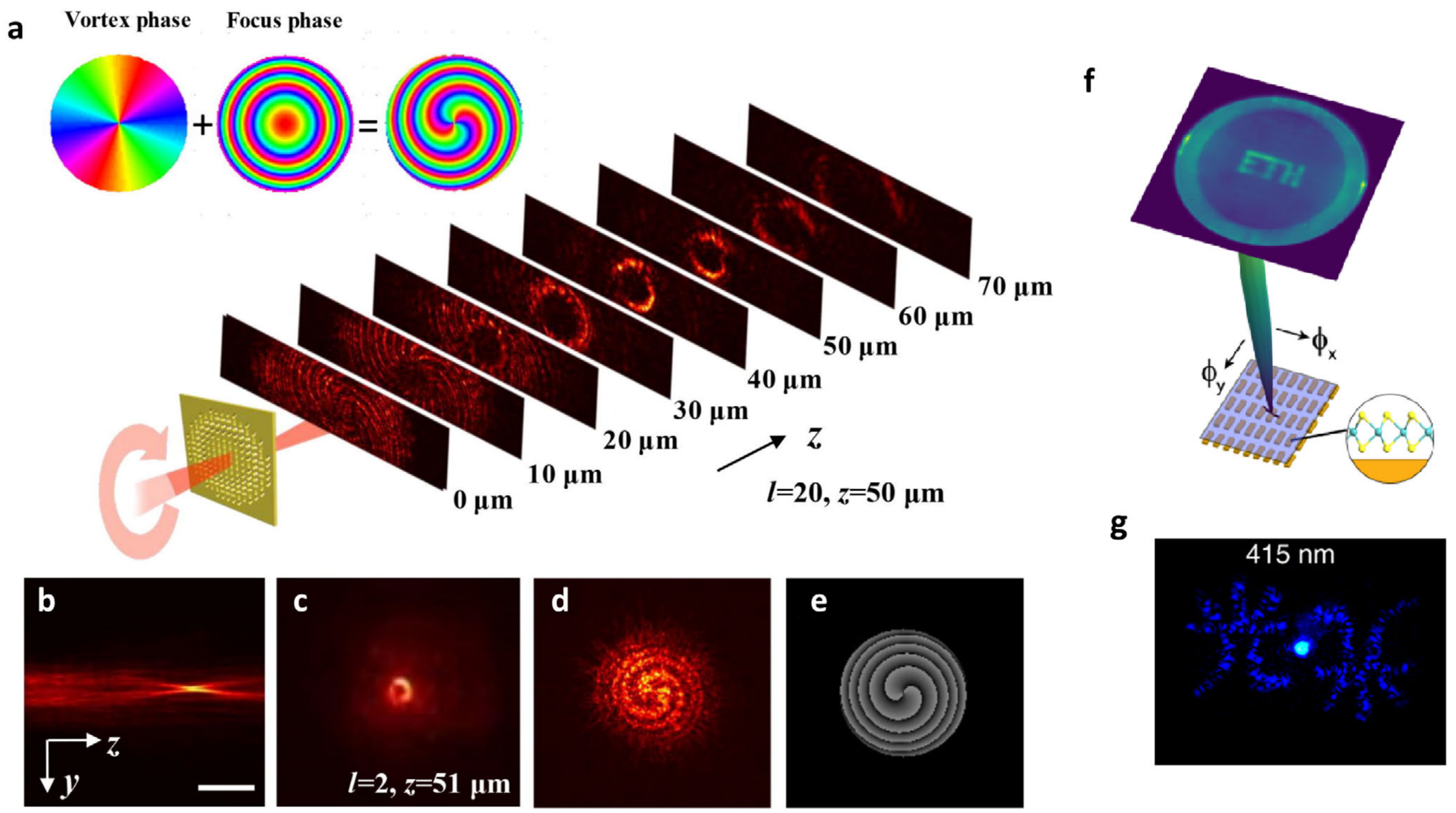

**Figure 11.** Light structuring. a–e) The focal vortex beams generation and coaxial interference detection. Emitted SH along the $z$-axis with topological charge $l$ equal to 20 (a). SH signal with $l$ = 2 cross section (b). SH signal field distribution in the focal plane (c). Interferometric image used to determine $l$, which was found to be 2 (d). The simulation results (e). Reproduced with permission.[252] Copyright 2020, Science China Press. Dynamic beam steering of SH emission of a $MoS_2$–gold phased array antenna (f). The phase delays $\phi_x$ and $\phi_y$ were varied in time by using a motorized lens system, tilting the wavefront for the exciting radiation, to write a "ETH" far-field pattern. Reproduced with permission.[256] Copyright 2019, American Chemical Society. Reconstructed holographic image at SH wavelength 415 nm (g). Reproduced with permission.[257] Copyright 2019, American Chemical Society.

As we discussed in Section 6, integration of TMDs with functional surfaces can be used to enhance the frequency conversion efficiency. However, integration of 2D materials on metasurfaces can also be used for light structuring. For instance, a desired Berry phase distribution can be induced by a gold nanohole array[250–253] or by nanorod antennas.[254] The emitted SH vortex beam inherits the OAM from the array due to the conservation of the angular momentum, allowing the experimental realization of high topological charges[252] (**Figure 11**a–e). Alternatively, a gold axicon metasurface coupled to a 1L TMD can be used to generate SH Bessel beams.[255]

A nanohole array can also be used to impart a spin-dependent Berry phase into the FF field. As a consequence, the nonlinear signal originating from different valleys of a 1L TMD integrated on such metasurface can be steered in a desired direction.[250] Three beams with different chirality are generated upon excitation with linearly polarized light, whereas only two beams are emitted when the metasurface is pumped with circularly polarized light. This pronounced effect is only possible due to the unique valley degree of freedom of TMDs.

Further, a controllable steering in a broad angle range of the harmonic signal beam from a 1L TMD can be achieved by integration with a phased array of gold antennas.[256] This result originates from the constructive interference of the signals coming from the individual antennas and the phase delays $\phi$ are achieved by laterally displacing the focus of the excitation beam with high precision (Figure 11f).

The possibility to control the phase of SH signals allows also nonlinear holographic imaging, which attracts great interest due to its ability to carry a large amount of information. Highly efficient holographic imaging at SH frequency was reported for nanopatterned 1L TMDs[257] (Figure 11g) and 1L TMDs covering a Au metasurface with arrays of nanoholes.[251] Zhao et al., further demonstrated that a precise control of the Berry phase can be exploited to perform chirality-selected SH holography.[253] As an alternative to the Berry phase modulation, an advanced wavefront manipulation can be achieved with an independent control the phase and amplitude of the SH beam by a resonant V-shaped nanohole array.[258]

## 8. Outlook and Conclusions

### 8.1. Perspectives and Open Challenges

One of the major open challenges for nonlinear optics with 2D materials is the enhancement of their absolute efficiency (e.g., the conversion efficiency for SHG and the nonlinear optical gain of OPAs, and frequency converters). Despite the huge intrinsic nonlinearity of monolayers and the absence of phase-matching constraints, the sub-nm propagation length still prevents optical amplifiers based on 2D materials from reaching high gain regimes, that is, comparable to the standards of bulk anisotropic nonlinear crystals such as beta-barium borate (BBO), bismuth borate (BIBO), lithium tetraborate (LBO), potassium titanyle arsenate (KTA), and potassium titanyl phosphate (KTP). Since the nonlinear optical properties of multi-layer TMDs critically depend on their crystallographic symmetry, we envision that $3R$ noncentrosymmetric TMDs might offer valuable opportunities in this direction, given their preserved second order nonlinear response also in the multi-layer form. On the other hand, upon increasing the thickness of the nonlinear crystal, phase-matching constraints start imposing limitations over the amplification bandwidth, in particular when the number of TMD layers exceeds $\approx 500$.[7] A further challenge in this direction is indeed the trade-off between the nonlinear gain and the amplification bandwidth. In this context, $3R$-TMDs possess huge nonlinearities (2–3 orders of magnitude higher than standard nonlinear crystals) and they could be used to reach regimes of high nonlinear optical gain at microscopic thickness ($1-10\mu$m) while retaining ultrabroad amplification bandwidths.

Beyond frequency converters, we can foresee the rise of two new fields of research related to nonlinear optics with layered materials. The first deals with twistronics:[259,260] a plethora of exotic electronic and optical properties emerges by stacking 2D materials at a defined twist angle, paving the way to new fundamental studies and photonic applications. The possibility to vary the efficiency of nonlinear processes in a broad spectral range by controlling the twist angle between the constituents of the layered heterostructure[261] provides a highly tunable platform for integrated nonlinear optical devices. Despite its attractiveness, to date only few works have studied the twist-dependent nonlinear optical properties of layered heterostructures,[83,261,262] and the potential of nonlinear twistronics has remained largely unexplored.[263] We can anticipate studies of twist-dependent NLO phenomena in bi- and few-layer graphene and TMDs related to moiré superlattices, which not only can give an additional knob to engineer the optical properties of layered materials, but are also responsible for the formation of new electronic states.[264] In this regard, NLO spectroscopy, and in particular near-field nonlinear imaging,[34] can shed light on the exotic properties of moiré excitons.

In addition, the use of layered materials for quantum optics and, in particular, for entangled-photon generation represents another highly promising research direction. Although entangled-photon pair generation is intrinsically limited in 2D materials by the small light–matter interaction volume,[265] the potential of TMD for SPDC[266] and spontaneous FWM represents an attractive research direction, and several groups all over the world are actively working on it. Specifically, time-energy entangled photons generated by SPDC possess unique temporal and spectral features, resulting in narrow distribution of the sum frequency of the entangled photon pair.[267] For photon energy entanglement, one can in fact generate a pair of entangled photons with very different energies provided that their sum matches the excitation photon energy.[268] Broadband entangled photons generation offers important advantages in quantum spectroscopy[268,269] and ghost imaging.[270] A wide entanglement bandwidth obviously requires the broadest possible phase-matching bandwidth and thus 2D materials offer a unique advantage in this respect. The demonstration of entangled-photon sources based on 2D materials will thus pave the way to a new generation of nanoscale and phase-matching-free quantum sources of light, also potentially integrable in compact optical circuits. Furthermore, on-chip entangled photon sources based on TMD crystals might also solve the long-lasting challenge of generating multi-photon entangled states while preserving a high degree of entanglement at the microchip level.[271] This might represent a significant step forward in the field of quantum information.

Finally, we anticipate further fundamental studies in 2D materials and related heterostructures thanks to the recent experimental and theoretical progress in the fast-developing field of solid-state HHG. The fact that the carrier motion (which drives the HHG process) in a solid is defined by its band structure, makes HHG a sensitive tool to probe the electronic properties of a system,[95] many-body effects,[272,273] Berry's curvature,[94] and topology.[274] We therefore anticipate further studies of 2D materials with HHG, including unconventional superconducting[275] and topological phases[276] reported in 1L TMDs. Beyond its applicability to study isolated materials, HHG is predicted to be sensitive also to the stacking order[277] and interlayer coupling,[277,278] which can also be utilized to manipulate the HHG properties (e.g., intensity and polarization). HHG can also play a key role in the aforementioned emerging field of twistronics: HHG in twisted heterostructures could be used to probe the electronic landscape of moiré superlattices and flat bands. Besides these unique possibilities to investigate elusive electronic properties of 2D materials, HHG spectroscopy also paves the way to control[279] and track (either with two-color HHG[280,281] or pump–HHG spectroscopy[282]) electronic dynamics at attosecond (sub-cycle) timescales. The progress in this field can thus lead to significant technological advances, such as nanoscale petahertz electronics and valleytronics, as well as to the understanding of a wide range of quantum phenomena in 2D materials.[283,284]

## 9. Conclusions

In this review, we highlighted the most recent advances in the growing and expanding field of parametric nonlinear optics with layered materials and related heterostructures, and we have outlined new conceptual and technological challenges, along with possible strategies to solve them. The 2D nature of graphene and TMDs, combined with their exceptionally high intrinsic nonlinearity, the ease of integration with other layered materials and with photonic platforms and their sensitivity to the surrounding environment, makes them exceptionally promising materials for both fundamental and applied nonlinear optics. We believe that the foreseeable future will bring new and exciting results in this field, in particular with respect to hybrid devices that combine layered materials with photonic platforms (e.g., metasurfaces, waveguides, fibers) or other low-dimensional systems (e.g., QDs, NWs). Finally, the strong nonlinearity of TMDs might also open new exciting pathways toward the generation of entangled-photon pairs at the nanoscale and thus to the realm of quantum technologies.

## Acknowledgements

The authors acknowledge the financial support from the European Union Horizon 2020 Programme under Grant Agreement No. 881603 Graphene Core 3. This publication is part of the METAFAST project that received funding from the European Union Horizon 2020 Research and Innovation programme under Grant Agreement No. 899673. G.S. acknowledges the German Research Foundation DFG (CRC 1375 NOA project number B5) and the Daimler und Benz foundation for financial support.

Open access funding enabled and organized by Projekt DEAL.

## Conflict of Interest

The authors declare no conflict of interest.

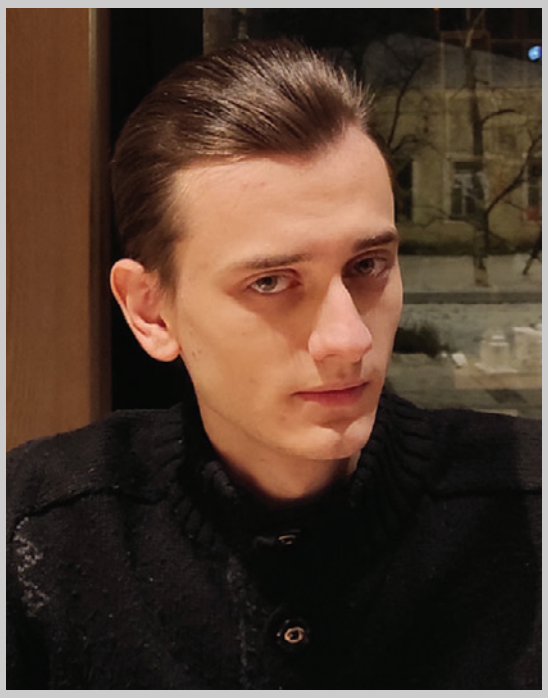

**Oleg Dogadov** is a Ph.D. student at Polytechnic University of Milan in the group led by Prof. Giulio Cerullo. He got his Master degree in applied mathematics and physics from Moscow Institute of Physics and Technology. During his studies he did research projects at the Photochemistry Center in Moscow (2016–2018) and at the Institute of Spectroscopy in Troitsk (2018–2020). In 2020 he was a visiting student in Ultrafast Spectroscopy group at EPFL in Lausanne. His research is focused on studying 2D and strongly correlated materials under nonequilibrium conditions using time-resolved spectroscopic techniques.

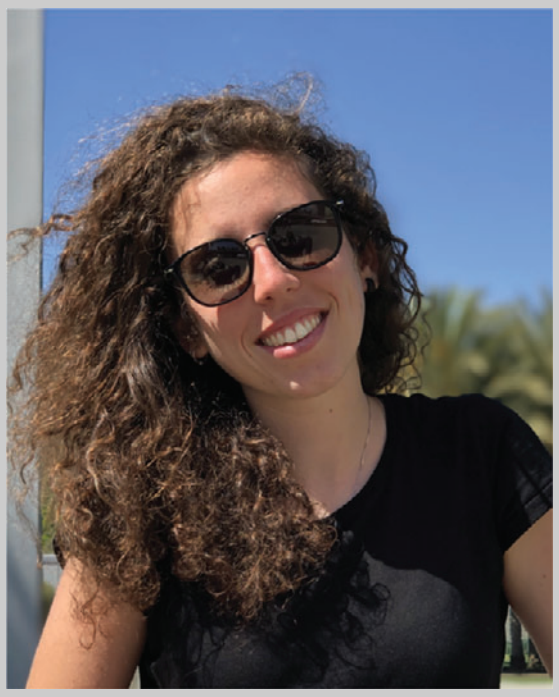

**Chiara Trovatello** is a postdoctoral research scientist at Columbia University. She did her Ph.D. in physics within the Graphene Flagship network under the supervision of Prof. Giulio Cerullo, and graduated in 2020. Her research activity focuses on ultrafast spectroscopy and nonlinear optics of low dimensional materials (graphene, TMDs, and van der Waals heterostructures). During her Ph.D. she visited the University of Würzburg in Germany (March 2018), and she was granted with a Marie Curie RISE Fellowship to visit Columbia University (March–September 2019). She is also Founder and Former President of the OPTICA Student Chapter of Milan.

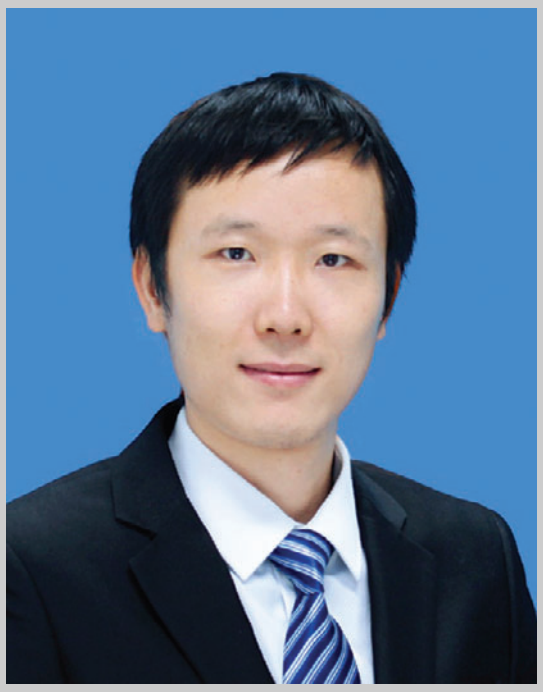

**Baicheng Yao** is a professor of information science at UESTC. He received his Ph.D. in 2016 from UESTC/UCLA (co-education), and worked as a research associate at University of Cambridge, 2017–2018. Focusing on optoelectronic and photonic devices, his work has appeared in 70+ journals and conferences (4 ESI highly-cited), including Nature, Nature Photonics, Nature Communications series amongst others. He won 10+ scientific and industrial honors and hold 20+ innovation patents. He serves as a deputy/associate editor of 3 journals, a reviewer of 20+ journals, and the co-chair/TPC-chair of 5 international conferences. He is a IEEE/OSA senior member.

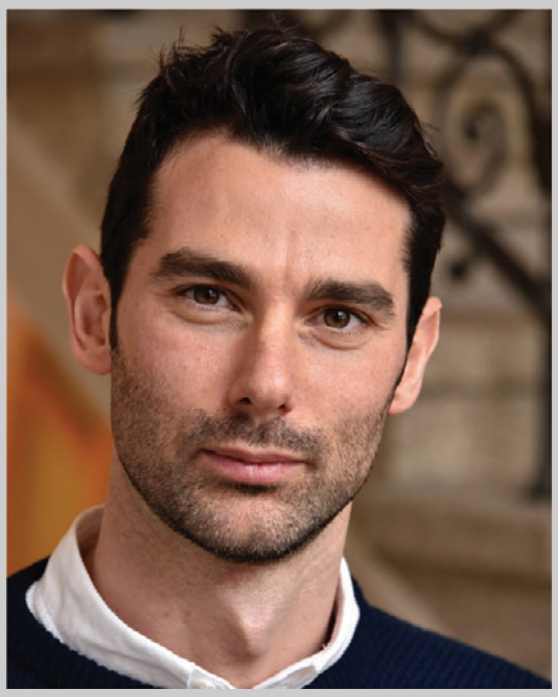

**Giancarlo Soavi** is a junior professor at the Institute of Solid State Physics of the Friedrich Schiller University in Jena. His research activity focuses on the ultrafast dynamics and nonlinear optical properties of low dimensional materials, in particular graphene and related layered materials. He obtained his Ph.D. in 2015 from Politecnico di Milano and he worked as a visiting scientist (2014) at the University of Konstanz (Germany) and as a research associate (2015–2018) at the Cambridge Graphene Centre (UK). He is the leader of the working group "Lasers and Sources" of the European Graphene Flagship.

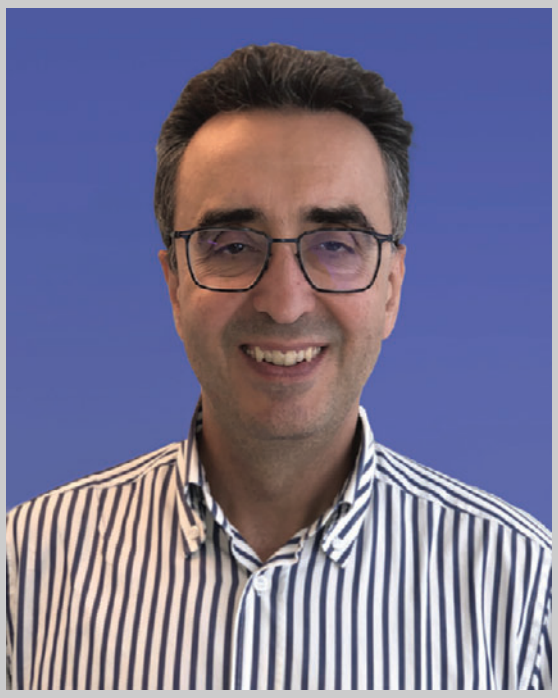

**Giulio Cerullo** is a full professor of physics at Polytechnic University of Milan. His research activity focuses on "Ultrafast Optical Science" and concerns the generation and manipulation of ultrashort light pulses, and their use to capture the ultrafast dynamics in molecules, nanostructures, and 2D materials. He is a Fellow of OPTICA and EPS, and Chair of the QEOD of EPS. He is on the Editorial Advisory Board of the journals Optica, Laser&Photonics Reviews, Scientific Reports, Chemical Physics, Journal of Raman spectroscopy and he is General Chair of CLEO-EU 2017, Ultrafast Phenomena 2018 and the International Conference on Raman Spectroscopy 2020.